\documentclass[showpacs,aps,prd,twocolumn,eqsecnum,floats,superscriptaddress]{revtex4}

\newif\ifpdf\ifx\pdfoutput\undefined\pdffalse\else\pdfoutput=1\pdftrue\fi
\usepackage{graphicx}
\usepackage{amsfonts}
\usepackage{amsmath}
\usepackage{subfigure}
\usepackage{bm}

\graphicspath{{./graphics/}}
\ifpdf\usepackage{hyperref}\else\fi
\ifpdf\DeclareGraphicsExtensions{.pdf,.jpg,.png}\else\fi

\def\Hz{\mathrm{Hz}}

\def\htilde{\tilde{h}(f)}
\def \st{\tilde{s}}

\def \A {{\cal A}}
\def \C {{\cal C}}
\def \tC {\tilde {\C}}
\def \V{{\cal V}}
\def \Nind {N_{\rm ind}}
\def \Nwin {N_{\rm win}}
\def \Ls {\Lambda^*}
\def \R {{\cal R}}
\def \e0{\epsilon_0}
\def \W {{\cal W}}
\def \h {{1 \over 2}}
\def \Dl {\Delta \lambda}
\def \Dt {\Delta \tau} 
\def \Rs {{\cal R}_{\cal S}}
\def \Rt {{\cal R}_{\cal T}}
\def\be{\begin{equation}}
\def\ee{\end{equation}}
\def\bea{\begin{eqnarray}}
\def\eea{\end{eqnarray}}
\def \no {\nonumber}
\def\lsim{\mathrel{\rlap{\lower4pt\hbox{\hskip1pt$\sim$}}
    \raise1pt\hbox{$<$}}}                % less than or approx. symbol
\def\gsim{\mathrel{\rlap{\lower4pt\hbox{\hskip1pt$\sim$}}
    \raise1pt\hbox{$>$}}}                % greater than or approx. symbol

\begin{document}

\preprint{IUCAA xx/2006}

\title{Detecting gravitational waves from inspiraling binaries with a network of detectors : coherent versus coincident strategies}

\author{Himan Mukhopadhyay}
%\email{himan@iucaa.ernet.in}
\affiliation{Inter-University Centre for Astronomy and Astrophysics,\\
Post Bag 4, Ganeshkhind, Pune 411007, India}

\author{Norichica Sago}
%\email{sago@vega.ess.sci.osaka-u.ac.jp}
\altaffiliation[Present address: ]{School of Mathematics, University of Southampton,
Southampton SO17 1BJ, United Kingdom}
\affiliation{Department of Earth and Space Science, 
Graduate School of Science, Osaka University, Toyonaka, 
Osaka 560-0043, Japan}

\author{Hideyuki Tagoshi} 
%\email{tagoshi@vega.ess.sci.osaka-u.ac.jp}
\affiliation{Department of Earth and Space Science, 
Graduate School of Science, Osaka University, Toyonaka, 
Osaka 560-0043, Japan}

\author{Sanjeev Dhurandhar}
%\email{sdh@iucaa.ernet.in}
\affiliation{Inter-University Centre for Astronomy and Astrophysics,\\
Post Bag 4, Ganeshkhind, Pune 411007, India}

\author{Hirotaka Takahashi}
%\email{hirotaka.takahashi@aei.mpg.de}
\affiliation{Max-Plank-Institut f\"{u}r Gravitationsphysik, 
Albert-Einstein-Institut, Am M\"{u}hlengerg 1, D-14476 Golm bei Potsdam, 
Germany}

\author{Nobuyuki Kanda}
%\email{kanda@sci.osaka-cu.ac.jp}
\affiliation{Department of Physics, Graduate School of Science, Osaka City University, 
Osaka 558-8585, Japan}

\date{August 22, 2006,  ver.9.7}

\begin{abstract}
We compare two strategies of multi-detector detection of compact binary inspiral signals, namely, the coincidence and the coherent. For simplicity we consider here two identical detectors having the same power spectral density of noise, that of initial LIGO, located in the same place and having the same orientation. We consider the cases of independent noise as well as that of correlated noise. The coincident strategy involves separately making two candidate event lists, one for each detector, and from these choosing those pairs of events from the two lists which lie within a suitable parameter window, which then are called as coincidence detections. The coherent strategy on the other hand involves combining the data phase coherently, so as to obtain a single network statistic which is then compared with a single threshold. Here we attempt to shed light on the question as to which strategy is better. We compare the performances of the two methods by plotting the \emph{receiver operating characteristics} (ROC) for the two strategies. Several of the results are obtained analytically in order to gain insight. Further we perform numerical simulations in order to determine certain parameters in the analytic formulae and thus obtain the final complete  results. We consider here several cases from the relatively simple to the astrophysically more relevant in order to establish our results.  The bottom line is that the coherent strategy although more computationally expensive in general than the coincidence strategy, is superior to the coincidence strategy - considerably less false dismissal probability for the same false alarm probability in the viable false alarm regime.

\end{abstract}
\pacs{95.85.Sz,04.80.Nn,07.05.Kf,95.55.Ym}

\maketitle
%%%%%%%%%%%%%%%%%%%%%%%%%%%%%%%%%%%%%%%%%%%%%%%%%%%%%%%%%%%%%%%%%%%%%%%%%%%
\section{Introduction}
\label{intro}

\par
The existence of gravitational waves (GW), predicted in the theory of
general relativity, has long been verified `indirectly' through the
observations of Hulse and Taylor. The inspiral of the
members of the binary pulsar system named after them has been successfully accounted for in
terms of the back-reaction due to the radiated GW  
\cite{HT}. However, detecting such waves with man-made `antennas' has not
been possible so far. Nevertheless, this problem has received a lot of
attention in the past several years, especially, due to the arrival of laser-interferometric
detectors, which are expected to have sensitivities close to that
required for detecting such waves \cite{GBGWD}. The space mission LISA 
\cite{lisa} is also planned by the NASA and ESA to detect low frequency GW. 
The significance of the direct detection of GW lies, not only
in the opening of an entirely new window into observational astronomy; 
it further promises to place our present theories of gravitation on an
entirely different foundation from that which currently exists.
Strong field gravity has never been directly observed and experimental
verification of the predictions of general relativity are very much
needed. GW detection will undoubtedly be the most exciting development in experimental general
relativity since the theory's birth in 1915.
\par
Because several detectors will take data simultaneously, it is advantageous to 
perform a multi-detector search for GW signals. A multi-detector search would 
(i) improve our confidence in detection of a GW event if a candidate event is 
registered; (ii) provide useful directional information on the GW event if the detectors 
are sufficiently separated in location, that is, at large geographical separations; 
(iii) provide polarization information if they are differently oriented.  
The information mentioned in (ii) and (iii) is degenerate in a single detector. 
\par
The response of a network to a GW is phase coherent. By `correcting' for the 
phase from the data of different detectors, and the time-delay between detectors 
(this amounts to bringing the detectors to the same site and with the same 
orientation), one can combine the data phase coherently. That this was achievable was 
shown in several works \cite{PDB}. Here the GW source was the well studied inspiraling 
compact binary.  
\par
In this paper we compare two different multi-detector detection 
strategies the coherent versus the coincident. The GW source we choose is the inspiraling 
binary because the phase of the wave can be sufficiently accurately computed; the 
phase of the wave has been modeled to the 3.5 post-Newtonian (PN) order \cite{arun} which 
means that the phase of the wave is accurate to about a cycle in a wave train of $\sim 10^4$ 
cycles for typical stellar mass binary inspirals. 
Secondly, in the context of GW detectors, it is of great astrophysical significance \cite{CBS}. 
Finn \cite{Finn} has performed similar analysis with a sinusoidal signal. 
However, our analysis goes further in that we consider 
a GW signal of astrophysical importance and our analysis uses template banks 
of inspirals to compare the two strategies.
\par   
The coherent search strategy uses the maximum likelihood method where  
a {\it single} likelihood ratio for the entire network is constructed - that is the 
network is treated as a single detector - this is similar to aperture 
synthesis carried out, for example, by radio astronomers. The 
likelihood method combines the data from a network of detectors in a 
{\it phase coherent} manner to yield a single statistic which is 
optimal in the maximum likelihood sense. This statistic is then compared with the 
threshold determined by the false alarm rate that we are prepared to 
tolerate. If the statistic crosses the threshold, then a detection is 
announced. Note that a {\it single} likelihood ratio is computed and 
a {\it single} threshold applied in this type of search.
\par
On the other hand, the coincidence approach involves separately 
filtering the signal in each detector, applying two separate thresholds 
corresponding to each detector and preparing two event lists determined 
by the crossings. Then the event lists are matched. If the estimated 
parameters for the events lie in a reasonable neighborhood in the 
parameter space of signals, a coincident detection is registered - 
if the differences in the estimated parameters lie within certain bounds - 
that is, the parameters of the events must lie within a certain `window' of the 
parameter space. A coincidence search for inspiraling binaries 
with real data has already been performed in many cases \cite{TAMA,LIGOS1S2,LIGOTAMA}. 
\par
The debate between the coherent and coincident strategies has existed in the community ever since 
the coherent strategy was formulated. While the coherent strategy involves more 
computational costs than the coincident strategy, we show in this paper for the 
simple case of two identical detectors in the same location and with same orientation 
that the performance of the coherent strategy is superior to that of the 
coincidence strategy - the false dismissal is considerably less for the coherent as compared  
with the coincident at the same false alarm. This 
is in spite of the fact that for simplicity we have taken the detectors to be co-aligned; 
we expect the coherent strategy to perform better in the case of non-aligned 
detectors than its competitor. In order to decide on the performance between the two 
strategies, we plot the false dismissal versus false alarm curves - the   
\emph{Receiver Operating Characteristic} (ROC) curves. We use the initial LIGO (LIGO I) noise curve 
\cite{ligoInoise} in our calculations and assume Gaussian stationary noise in our simulations.     
\par
The paper is organized as follows: In section II we briefly review the signal in a 
network of detectors, the post-Newtonian inspiral waveform and also the maximum likelihood 
analysis. In section III we obtain analytical results of false alarm and false 
dismissal probabilities for both coherent and coincident strategies. In this section 
we also consider the case of correlated noise between the detectors. This case would 
be of particular significance to the two LIGO detectors at Hanford and other similar 
future detector topologies elsewhere envisaged. In section IV we perform simulations. The simulations are performed in order(i) to determine the number of independent templates and, (ii) to 
determine the size of the parameter window for the coincidence case. Simulations are necessary because at 
low SNR $\sim 10$, the Fisher information matrix gives poor estimates of the errors in parameters and therefore grossly undersizes the window. Further, simulations are also performed in order to plot the ROC curves which validate our analytical results. In section V we summarize the results and describe future directions.      

%%%%%%%%%%%%%%%%%%%%%%%%%%%%%%%%%%%%%%%%%%%%%%%%%%%%%%%%%%%%%%%%%%%%%%%%%%
%%%%%%%%%%%%%%%%%%%%%%%%%%%%%%%%%%%%%%%%%%%%%%%%%%%%%%%%%%%%%%%%%%%%%%%%%%%
\section{Network Signal}
\label{primer}

%%%%%%%%%%%%%%%%%%%%%%%%%%%%%%%%%%%%%%%%%%%%%%%%%%%%%%%%%%%%%%%%%%%%%%%%%%%%%%%%%%%
\subsection{The signal}
\label{signal}
In this section, we describe the necessary background and notation in order to make the 
paper self contained. Generic time domain functions $h(t)$ will be denoted by $\htilde$ in the Fourier domain, where,
\begin{equation}
\htilde = \int_{-\infty}^{\infty}dt\ h(t) e^{2\pi ift}.
\label{eq:fftdef}
\end{equation}
Although engineering and much of LIGO analysis software uses the opposite sign for the exponent in (\ref{eq:fftdef}), we maintain this notation for consistency with the published literature \cite{tanaka}.
\par
In order to describe the response of the detectors to the wave we need to define the relevant quantities that appear in the response. Because a lot of our results are obtained numerically from simulations we give a somewhat detailed description here. The dependence of the response on the angles is best given by defining two frames:  
\par
{\it Wave frame $(X,Y,Z)$}: The gravitational wave travels along the positive $Z$ direction and $X$ and 
$Y$ denote the axes of polarization in such a way that a right-handed coordinate system is formed.
\par
{\it Detector frame $(x, y, z)$} : This denotes the orthogonal coordinate frame
attached to the detector. The arms of the detectors lie in the $x - y$ plane,
which is the plane tangent to the surface of the Earth 
and the $x$ axis bisecting the angle between the two arms and $y$ is chosen such that
the frame forms a right-handed coordinate system with the $z$ axis pointing radially 
out of the surface of the Earth. We just need one such frame because the detectors have the same 
orientation. 
\par
The Euler angles $(\phi, \theta, \psi)$ rotate the detector frame to the wave frame where we have chosen the Goldstein \cite{Goldstein} convention. Also let $\iota$ denote the angle between the line of sight and the orbital angular momentum vector of the binary. We directly write down the signal response 
$\htilde$ in the Fourier domain. In the stationary phase approximation the spin-less, restricted post-Newtonian 
inspiral signal is given by, 
\bea
\no
\htilde &=& {\mathcal{N}} \times E (\phi, \theta, \psi; \iota) f^{-7/6} \\
&& \times \exp ~ i \Psi(f; t_c, \delta_c, \tau_0, \tau_3) ,
\label{eq:stilde}			   
\eea
where $t_c$ and $\delta_c$ are respectively the coalescence time and the coalescence phase of the binary.
$E(\phi, \theta, \psi; \iota)$ is the extended antenna pattern function \cite{PDB} which depends on the orientation angles - 4 angles in this case - and is given by:
\be
E = \left [ \left ( \frac{1 + \cos^2 \iota}{2} \right )^2 F_+^2 (\phi, \theta, \psi) + \cos^2 \iota F_{\times}^2 (\phi, \theta, \psi) \right ]^{1/2}
\label{exantptn} 
\ee
where the $F_{+, \times}$ are the usual antenna pattern functions:
\bea
\no
F_+ &=& - \sin 2 \phi \frac{1 + \cos^2 \theta}{2} \cos 2 \psi - \cos 2 \phi \cos \theta \sin 2 \psi , \\
\no
F_{\times} &=& - \cos 2 \phi \cos \theta \cos 2 \psi + \sin 2 \phi \frac{1 + \cos^2 \theta}{2} \sin 2 \psi . \\ 
\label {antptn}
\eea
 
The factor $\mathcal{N}$ depends on $m_1, m_2$ the individual masses of the binary and the 
distance $r$ to the binaries in the following way:
\be 
{\cal N} = \left(\frac{5}{24}\right)^{1/2}
\frac{M^{5/6}\eta^{1/2}\pi^{-2/3}}{r} \, ,  
\label{N}
\ee 
where we have resorted to geometrized units of $c = G = 1$ for convenience. 
$M = m_1 + m_2$ is the total mass of the binary system and 
$\eta$  $=m_1 m_2 / M^2$ is the ratio of reduced mass to the total mass $M$. 
To ease the computation, a new set of time parameters $\{ \tau_0, \tau_3 \}$
that are functions of the masses is chosen such that the template spacing is approximately uniform over the parameter space in these parameters. 
These parameters are defined as: 
\bea
\no \tau_0 &=& \frac{5}{256 \pi \eta f_a} \left ( \pi M f_a \right )^{-5/3}, \\  
\tau_3 &=& \frac{1}{8 \eta f_a} \left ( \pi M f_a \right )^{-2/3},
\eea
where $f_a$ is a fiducial frequency usually chosen to be the seismic 
cut-off frequency. In this paper we will choose this frequency to be $40$ Hz.

The function $\Psi(f; t_c, \delta_c, \tau_0, \tau_3)$ describes the phase evolution 
of the inspiral waveform. We adopt the 3PN formula given by 
\begin{eqnarray}
\nonumber
&& \! \! \! \! \! \! \! \! \! \Psi(f;t_c,\delta_c,\tau_0,\tau_3)  = 2 \pi f t_c - \delta_c - \frac{\pi}{4}\\
&& \! \! \! \! \! \! \! \! \! \! \! \!+ \frac {3}{128 \eta} (\pi M f)^{-5/3} \sum_{k=0}^{6} \alpha_k (\pi M f)^{k/3} \, ,
\label{phase}
\end{eqnarray}
where the coefficients are as follows:
\begin{eqnarray}
\no \alpha_0 &=& 1, \\
\no \alpha_1 &=& 0, \\
\no \alpha_2 &=& \frac{20}{9} \left( \frac{743}{336} + \frac{11}{4} \eta \right), \\
\no \alpha_3 &=& -16 \pi, \\
\no \alpha_4 &=& 10 \left( \frac{3058673}{1016064} + \frac{5429}{1008} \eta + \frac{617}{144} \eta^2 \right), \\
\nonumber
\alpha_5 &=& \pi \left( \frac{38645}{756} - \frac{65}{9} \eta \right) \left(1 + \ln \left(\pi M f\right) \right),\\
\nonumber
\alpha_6 &=& \left( \frac{11583231236531}{4694215680} - \frac{640 \pi^2}{3} - \frac{6848 \gamma}{21} \right)\\
\nonumber
& + &  \eta \left( -\frac{15335597827}{3048192} + \frac{2255 \pi^2}{12} - \frac{1760 \theta}{3} \right.\\
\nonumber
& + & \left. \frac{12320 \lambda}{9} \right)  +  \frac{76055}{1728} \eta^2 - \frac{127825}{1296} \eta^3\\
\no & - & \frac{6848}{21} \ln \left[4 ( \pi M f )^{1/3} \right]\, . \\
%\nonumber
%\alpha_7 & =& \pi \left( \frac{77096675}{254016} + \frac{378515}{1512} \eta 
%- \frac{74045}{756} \eta^2 \right) \, . \\
\end{eqnarray}
The quantity $\gamma$ is the Euler constant, $\lambda = -1987/3080$ 
and $\theta = -11831/9240$.  
\par
Although the highest PN order obtained to date  for a two point mass inspiral
is 3.5 PN order,  for a stellar mass binary system 
in the typical frequency bandwidth of ground-based detectors from few tens 
of Hz to kHz, the 3 PN waveform is accurate to within a cycle in about a total 
of $10^4$ cycles.  

\subsection {Matched filtering and normalized templates}

It is important to distinguish between the nature of parameters that appear in the above {\em form} of
the GW chirp. The two mass parameters $\vec \mu \equiv \{ M,\eta \}$ or equivalently $\tau_0, \tau_3$ are the dynamical parameters and determine the {\em
shape} of the chirp. Exclusive to this set are the `offset' parameters $\vec \lambda
\equiv \{ t_c,\delta_c \}$ which determine the duration of the chirp or the end-points in the time series. The latter can be quickly estimated in
the matched-filtering paradigm without having to construct template banks over them. For example,  
spectral correlators using FFTs allows us to estimate correlations at all time lags and thus estimate 
$t_c$. Similarly, using the quadrature formalism to construct template banks, the coalescence phase can be 
estimated analytically. However, the dynamical parameters need to be searched over by a grid of 
templates spanning the parameter space; thus this is the effective parameter space. The template bank is set up over $\vec \mu \equiv \{ \tau_0,\tau_3 \}$. The form of (\ref{eq:stilde}) allows us to write the explicit quadrature representation of the $i$-th template as, 
\bea
\no
\tilde{h}(f;\vec \mu_i, t_c, \delta_c) &=& \A (\st_0 (f; \vec \mu_i, t_c) \cos \delta_c \\    
                                    &+&  \st_{\pi/2}(f; \vec \mu_i, t_c) \sin \delta_c) \, ,
\label{eq:quadrature}
\eea
where,
\be
\st_0 (f; \vec \mu_i, t_c) = i \st_{\pi/2}(f; \vec \mu_i, t_c) \,.
\label{eq:sopi} 
\ee
The quantity $\A$ in (\ref{eq:quadrature}) is the amplitude of the waveform $h$. This is conveniently seen by demanding that both the templates $s_0$ and $s_{\pi/2}$ have have unit norm; 
i.e. the scalar products $(s_0,s_0) = (s_{\pi/2}, s_{\pi/2})  = 1$. The scalar product $(a,b)$ of two real functions $a(t)$ and $b(t)$ is defined as 
\be
\left ( a, b \right ) = 2 \int_{f_l}^{f_u} df \ \frac{  \tilde{a}(f) \tilde{b}^*(f) \ + \ 
\tilde{a}^*(f) \tilde{b}(f)}{S_h(f)},
\label{eq:scalar}
\ee
where, we use the Hermitian property of Fourier transforms of real functions. $S_h(f)$ is the one
sided power spectral density (PSD) of the noise of initial LIGO detectors \cite{ligoInoise}. Further, 
$f_l \leq f \leq f_u$ is taken to be the effective spectral window for computing the scalar product. 
The lower frequency cut-off $f_l$ is dependent on the sensitivity of the detector to seismic 
vibrations and is taken to be $40\ \Hz \equiv f_a$,  whereas the upper frequency cut-off $f_u$ is equal to half the full sampling frequency taken to be $2048\ \Hz$. In practice, the upper cut-off frequency 
for normalizing the templates, we choose to be 1 kHz, that is, we truncate the waveform at 1 kHz. The truncation frequency should be chosen to be the minimum of $f_{lso}$ and 1 kHz, but for the masses that we choose in our simulations,  $m_1, m_2 \sim 1.4 M_{\odot}$, the $f_{lso} \sim 1600$ Hz which is larger than 1 kHz. 
\par
From the definition of the scalar product Eq. (\ref{eq:scalar}) we deduce from Eq. (\ref{eq:sopi})  that $s_0$ and $s_{\pi/2}$ are orthogonal, i.e. $(s_0, s_{\pi/2}) = 0$. Also from Eq. (\ref{eq:quadrature}) we see that $(h, h) = \A^2$.  
\par
Using the normalized templates a template bank is generated which covers the parameter space with a given minimal match usually chosen to be about 0.97 (See \cite{owen}); the correlation of a signal with a template must not fall below the specified minimal match fraction for at least one template in the template bank. The coarsest bank satisfying these conditions is constructed. Having thus set up a template bank, the statistic $\rho$ for an output signal $h(t)$
of the interferometer is the maximum of the signal to noise ratio (SNR) over all the templates (labeled 
by $i$) and the time-lags labeled by $t_c$. Thus, the statistic $\rho$ is given \footnote{In  much of recent literature as also in the LIGO Algorithms Library documentation \cite{lal}, the square root in
(\ref{eq:rho}) is not taken.} by,
\be
\rho = \max_{i, t_c} \left [(s_0, h)^2 + (s_{\pi/2}, h)^2 \right ]^{\h},
\label{eq:rho}
\ee
which is then compared with a pre-determined threshold. Any crossing above this threshold is recorded as a candidate event. A detailed description of this procedure can be found in \cite{lal,grasp,monty1,sd1,sd2}.

\subsection {Complex templates and correlations}

In view of the fact that we require just two templates for $\delta_c = 0, \pi/2$, namely, $s_0$ and 
$s_{\pi/2}$, it is found useful to combine the two templates into a single complex normalized template $S$. We then obtain complex correlations by taking scalar product with the data from each detector. In particular, the coherent statistic is constituted from these complex correlations \cite{PDB}. In the Fourier domain the complex template is given by:
\begin{equation}
%\tilde{S}(f) = \frac{2}{g} \sqrt{ \frac{2}{3 f_{a}}} \left( \frac{f}{f_{a}} \right)^{-7/6} 
%\hbox{exp}
%\left[~ i \Psi (f;t_c,\delta_c = 0,\tau_0,\tau_3)\right].
\tilde{S}(f) = \frac{1}{g} f^{-7/6}
\hbox{exp}\left[~ i \Psi (f;t_c,\delta_c = 0,\tau_0,\tau_3)\right].
\end{equation}
The normalization factor $g$ is chosen so that $(S,S) = 2$; it is given by, 
\be
%g^2 = \frac{4}{3} f_{a}^{4/3} \int_{f_{a}}^\infty \frac{df} {f^{7/3} S_{h}(f)} \, .
g^2=2 \int_{f_{a}}^\infty \frac{df} {f^{7/3} S_{h}(f)} \, .
\ee
In terms of $g$, the amplitude of the waveform $h$ is 
%$\A = \sqrt{3 f_a} g \mathcal{N} E / 2 $.
$\A = \mathcal{N} E g /\sqrt{2}$. 
\par 
For the two detector network, the data consists of two data trains, $\{ x^I(t) | I=1,2; 
\hbox{and} \; t \in [0,T] \}$ where data is taken in the time interval $[0, T]$. Assuming additive noise $n^I$ in each detector we have:
\be
x^I = h + n^I , ~~~ I = 1, 2 
\ee
Note that the signal is the same in each detector because we have assumed identical detectors in the same location and with the same orientation. The noise is however different in each detector. But since we assume identical noise PSDs and also assumed stationarity of the noise, the noise random variables satisfy the statistical property:
\be
\langle n^I (f) n^{*}_I(f') \rangle = \h  S_h (f) \delta (f - f') \,, 
\label{psd}
\ee
where, the angular brackets denote ensemble average.
\par
If the detectors have uncorrelated noise, $\langle n^1 (f) n^{*}_2(f') \rangle = 0$. On the other hand, if the detectors have correlated noise, then we may assume:
\be
\langle n^1 (f) n^{*}_2(f') \rangle = \h \epsilon (f)  S_h (f) \delta (f - f') \, ,
\label{noisecorr}
\ee
where $\epsilon$ is real and frequency dependent. In the time domain, we assume $\epsilon$ as a function of $|t - t'|$. 
This assumption implies also stationarity in the correlated noise. 
The quantity $\epsilon$ is, in general, a function of frequency, because different physical mechanisms generating noise are operative in different
frequency regimes. Consequently, the noise correlation is frequency dependent. 
In reality, there are several complications: the noise generated by a particular mechanism may drift in frequency with time and/or the $\epsilon$ may be dependent on  
the amplitude of noise, because of non-linear effects, etc. It is not easy to deal with such generic cases. However, when the correlation of noise is not very large, 
it is expected that the simple correlation model (\ref{noisecorr}) we adopt is adequate.
\par
If $\epsilon = 0$, the noise in the two detectors is uncorrelated; on the other hand if $\epsilon = 1$ then the noise is perfectly correlated. The cases of physical interest lie in between these two extremes.  However, later we will show that we can treat $\epsilon$ effectively as constant, thus simplifying the analysis. In the following, we analyze both cases of uncorrelated as well as correlated noise. 
\par
We end this section by defining the complex correlations $C^I$ which are particularly required in the coherent detection strategy. We retain the notation and definition as in \cite{PDB}. The complex conjugate of $C^I$ is denoted by $C_I^*$. It is given by:    
\begin{equation}
C_I^* = (S, x^I) = c_0^I - i c_{\pi/2}^I \,,
\end{equation}
where, $c_0^I$ and $c_{\pi/2}^I$ are the real and imaginary parts of $C^I$; they are obtained by taking the scalar products of the data $x^I$ with $s_0$ and $s_{\pi/2}$ respectively; that is,
\be
c^I_{0, \pi/2} = (s_{0, \pi/2}, x^I) \,.
\ee 

\section{Coherent, Coincident strategies and false alarm, false dismissal rates}
\label{strat}

This section serves as a prelude to the next section, section IV, where we obtain the ROC curves for each strategy. As mentioned before, these curves are obtained by plotting the false dismissal rate (or equivalently the detection probability) versus the false alarm probability for a given strategy. A fair comparison of the strategies is obtained by plotting the ROC curves for each such strategy. We consider the following four cases:

\begin{description}
\item[A. 1.] Coherent detection and uncorrelated noise, 
\item[A. 2.] Coherent detection and correlated noise,
\item[B. 1.] Coincident detection and uncorrelated noise,
\item[B. 2.] Coincident detection and correlated noise.
\end{description}

A comparison of the performance of the strategies is drawn pairwise between cases A. 1. and B. 1, and A. 2 and B. 2.
\par
In this section we obtain useful analytic formulae for the false alarm and false dismissal probabilities. We find that the probabilities are obtained in terms of certain parameters, the number of independent templates $\Nind$ and the size of the parameter window $\Nwin$, which are difficult to compute  analytically - we elaborate on these issues later in the text. However these quantities can be determined numerically via simulations. We pin down these parameters later in  section IV so that the relevant probabilities are fully determined. Only then the ROC curves can be plotted for each strategy.
     
\subsection{Coherent detection}
\label{coh}

In the coherent strategy, the data from different detectors are combined phase coherently so that the network operates effectively as a single detector. The maximum likelihood network statistic for coincident and co-aligned detectors has already been worked out in \cite{PDB} for the case of uncorrelated noise. The maximum likelihood statistics incorporates automatically the phase of the signal and results in the coherent statistic. The maximum likelihood approach tends to outperform any other adhoc test and we follow this approach for the case at hand. A detailed discussion on this issue, pertaining to aligned and misaligned detectors has been investigated by \cite{KMRM}. 

\subsubsection{Uncorrelated noise}
\label{uncorr}

The coherent network statistics for this case is given by:

\begin{equation}
L = \frac{1}{\sqrt{2}} |C_1^* + C_2^*| \,,
\label{stat1} 
\end{equation}
where the $C^I$ are the complex correlations of the two detectors $I = 1,2$. 
Although the statistic $L$ has the advantage that it is proportional to the amplitude of the GW, the square of $L$ has simpler form of the probability distribution if the noise in the detectors is Gaussian and stationary. Thus we define: 
\begin{equation}
\Lambda = L^2 = \frac{1}{2} \left[ \left(\sum_{I=1}^{2} c_{0}^I\right)^2 + \left(\sum_{I=1}^{2} c^I_{\pi/2}\right)^2 \right].
\label{stat2}
\end{equation}
Equivalently, $\Lambda$ can also be used to decide the presence or absence of a signal. Our goal is to obtain the false alarm and detection probabilities $P_{FA}$ and $P_{DE}$ respectively for the detection strategy. The false alarm probability is obtained in two steps: we obtain the probability  in the single template case (for instance, when the signal parameters are known) and then deduce the probability where one must search over the parameter space.
\par
Consider first the single template case:
\par    
Note that when the noise is uncorrelated, $c_0^I$ and $c_{\pi/2}^I,~ I = 1, 2$ are mutually independent 
Gaussian variables distributed standard normal; that each have mean zero and standard deviation unity. This is when there is no signal present in the data. From this it follows that the quantities 
$\frac{1}{\sqrt{2}}(c_{0}^1 + c_{0}^2)$ and $\frac{1}{\sqrt{2}}(c_{\pi/2}^1 + c_{\pi/2}^2)$ are also two independent Gaussian random variables with zero mean and unit variance.  Then $\Lambda$ which is the sum of squares of these two random variables has the probability density function (pdf) given by:
\begin{equation}
p_0(\Lambda) = \frac{1}{2} \hbox{exp} \left( - \frac{\Lambda}{2} \right).
\label{FA1}
\end{equation}
\par
Given a threshold $\Lambda^*$, the statistic $\Lambda$ exceeds the threshold $\Lambda^*$ with the probability:
\begin{equation}
P_{FA}^{\rm 1~template} = \int_{\Lambda^*}^\infty d\Lambda p_0(\Lambda) = \hbox{exp} \left( - \frac{\Lambda^*}{2} \right).
\end{equation}
 $P_{FA}^{\rm 1~template}$ is the false alarm probability for one template. Normally the threshold $\Lambda^*$ is chosen sufficiently high so that few false alarms occur during the observation period.
\par  
However, the parameters of the signal are not known a priori and one must scan the data with a bank of  templates. Filtering the data through all the templates in the bank enormously increases the false alarm rate. Typically, the templates densely cover the parameter space in order to minimize the chance of a signal being missed out. We have already briefly described such a template bank in section II. B. For the spinless inspiral the template bank comprises of templates in the parameter space spanned by the parameters $\{\tau_0, \tau_3\}$. However, when determining the false alarm rate, the parameter $t_c$ is as important, although one uses the Fast Fourier Transform (FFT) algorithm to efficiently scan over the parameter $t_c$ and excludes it from the standard template bank; one effectively also has templates in 
$t_c$ contributing to the false alarm rate.  If the number of templates in the three parameter extended bank (which includes $t_c$ also as a parameter besides $\tau_0, \tau_3$) are $N_{\rm bank}$ and if they all produced  independent filtered outputs and if the false alarm probability for a single template is small, then the false alarm rate for the extended bank is just that of the single template multiplied by the factor $N_{\rm bank}$. However, one finds that, because the templates are closely packed, the filtered outputs are strongly correlated. The dense coverage of templates results both from the high sampling rate in $t_c$ as well as the high minimal match chosen. Because of the correlations the false alarms are substantially reduced. It has been shown \cite{Rice} that the effect of the correlations is to effectively reduce the number of independent random variables. We find in section IV from simulations that this effective number of independent random variables $\Nind$ is much less than $N_{\rm bank}$. Thus the false alarm  probability for the template bank takes the form:
\begin{equation}
P_{FA} = \Nind \hbox{exp} \left( - \frac{\Lambda^*}{2} \right) \,.
\label{ndetecfa}
\end{equation}
\par
When a signal $h$ with amplitude $\A$ is present in the data $x^I$, the means of the correlations $c^I_0, c^I_{\pi/2}$ are no more zero, although they are still Gaussian distributed with variance unity. Now the pdf of $\Lambda$ given by Eq. (\ref{stat2}) is off-centered and is given by \cite{Hel, PDB}:
\begin{equation}
p_1(\Lambda)  = \frac{1}{2} \hbox{exp} \left[-\frac{\Lambda + (\sqrt{2} \A)^2}{2} \right] I_0 [(\sqrt{2} \A) \sqrt{\Lambda}] \,,
\label{detpr}
\end{equation}
where, $I_0$ is the modified Bessel function of the first kind of order zero. For large amplitudes $\A >> 1$, transforming the distribution in $\Lambda$ to one in $L$, one obtains approximately Gaussian distribution with mean equal to $\sqrt{2}$ times $\A$. This is the familiar 
$\sqrt{N}$ factor of enhancement one obtains in the signal-to-noise ratio (SNR) with $N$ independent detectors. (In fact the relevant factor 2 in Eqs.(\ref{stat2}, \ref{detpr}) is changed to $N$ in the case of a network of $N$ independent co-aligned and co-located detectors). Since we have assumed a network of two detectors with uncorrelated noise this factor is just $\sqrt{2}$.  
\par
By integrating $p_1(\Lambda)$ in Eq. (\ref{detpr}) we can easily obtain the false dismissal or the detection probability. Given a threshold $\Lambda^*$, the false dismissal probability 
is given by
\begin{equation}
P_{FD} = \int_0^{\Lambda^*} d\Lambda p_1(\Lambda). 
\end{equation}
The detection probability $P_{DE}$ is just $1 - P_{FD}$.

\subsubsection{Correlated noise}
\label{corr}

We now consider the case when the noise in the two detectors is correlated. For instance, the LIGO detectors at Hanford are in the same location and also co-aligned and therefore may share some of the noise sources; the same physical mechanisms would introduce noise in the two detectors giving rise to correlated noise.  The future planned detector, LCGT \cite{LCGT}, is designed such that two interferometers with the same length are installed in the same vacuum system. 
We thus expect some amount of correlation in noise.  
Here we consider the two detectors which are co-aligned and co-located with identical PSDs and Gaussian stationary noise.
We assume that the noise in the two detectors satisfies the properties given by Eqs. (\ref{psd}, \ref{noisecorr}). Then the quantities $c_{0,\pi/2}^I , \, I=1,2$ are also Gaussian random variables. 
\par
We define a complex vector whose components are the complex correlations $C^I$:
\begin{equation}
\tC^T=(C^1,C^2) \,.
\end{equation}
The superscript `T' denotes the transpose of a matrix. 
The covariance matrix of $C_I$ is given by
\begin{equation}
\tilde{\R}\equiv \langle \tC \tC^\dag\rangle =
\left(
\begin{array}{cc}
2 & 2\e0 \\
2\e0 & 2
\end{array}
\right) \,,
\label{covtilde}
\end{equation}
where $\tC^\dag=(C_1^*,C_2^*)$. 
The quantity $\e0$ is a weighted `average' of the noise correlation $\epsilon (f)$ and is given by:
\be
\e0 = 4 \int_{f_l}^{f_u} df \frac {\epsilon(f) |\tilde{s}_0 (f)|^2}{S_h (f)} \,.
\ee
The off-diagonal terms in $\tilde{\R}$ imply correlated data in the two detectors. Statistically independent data streams are obtained by forming linear combinations of the data streams from each detector. This procedure amounts to diagonalizing $\tilde{\R}$. Analogous methods were followed in \cite{Finn}. We therefore  obtain the eigenvalues and eigenvectors of $\tilde{\R}$. The eigenvalues are $2(1 \pm \e0)$. 
We choose two eigenvectors defined by
\bea
\nonumber 
(\V^+)^T &=& (1, 1)/\sqrt{2} \,, \\ 
(\V^-)^T &=& (1, -1)/\sqrt{2} \,,
\label{corrvec} 
\eea
with eigenvalues $2(1 + \e0)$ and $2(1 - \e0)$ respectively. 
We then define a orthogonal matrix $U$  by
$U=(\V^+ \V^-)$ which diagonalizes $\tilde{\R}$.  We define a vector 
$\C'=U\C=(C^+, C^-)^T$ where
\be
C^{\pm}  = \frac{1}{\sqrt{2}}(C^1 \pm C^2) \,.
\ee
The new complex correlations, $C^{\pm}$, represent the so called pseudo-detectors. 
%and this choice of eigenvectors allows us to frame the coherent statistic in terms of the complex correlations $C^I$. The eigenvectors given in Eq. (\ref{corrvec}) imply that the relevant linear combinations of the data are essentially just the sum and difference of the data from the two detectors. Accordingly, we form the statistics:
%\be
%C^{\pm}  = \frac{1}{\sqrt{2}}(C^1 \pm C^2) \,.
%\ee
The off-diagonal terms in their covariance matrix are zero. Thus, we can 
treat them as independent variables. 
We also observe that the mean of $C^-$ is zero even in the presence of a signal. This suggests that we need to consider only $C^+$ to build our detection statistic. Note that $C^+$ is the output of a matched filter and thus is optimal. Using the covariance matrix $\tilde{\R}$, it is easy to verify that the real and imaginary parts of $C^+$ are Gaussian distributed with variance $1 + \e0$. In the absence of a signal $\langle C^1 \rangle = \langle C^2 \rangle = 0$ which implies $\langle C^+ \rangle = 0$. Defining the coherent detection statistic as $\Lambda = |C^+|^2$ we obtain the distributions for 
$\Lambda$ in the absence and presence of a signal. In absence of a signal we have: 
\begin{equation}
p_0(\Lambda) = \frac{1}{2(1+\e0)} \exp \left( - \frac{\Lambda}{2(1+\e0)} \right) \,,
\end{equation}
while in the presence of a signal with amplitude $\A$:
\bea
\nonumber
p_1(\Lambda) &=& \frac{1}{2(1+\e0)} \hbox{exp} \left[-\frac{(\Lambda + 2 \A^2)}{2(1+\e0)} \right] \\
             &\times& I_0 \left (\A \frac{\sqrt{2 \Lambda}}{(1+\e0)} \right).
\eea
From these probability densities we find the false alarm and detection probabilities. The false alarm probability when the data is passed through $\Nind$ independent templates is given by: 
\bea
\nonumber
P_{FA} (\Ls) &=& \Nind \int_{\Lambda^*}^\infty d\Lambda p_0(\Lambda) \\ 
       &=& \Nind \exp \left( - \frac{\Lambda^*}{2(1+\e0)} \right),
\label{pfacohc}
\eea
and the detection probability is given by:
\begin{eqnarray}
\nonumber
P_{DE} (\Ls) & = & \int_{\Lambda^*}^{\infty} \frac{d\Lambda}{2(1+\e0)} 
\hbox{exp} \left[-\frac{(\Lambda + 2 \A^2)}{2(1+\e0)} \right]\\
& \times & I_0 \left (\A \frac{\sqrt{2\Lambda}}{(1+\e0)} \right).
\end{eqnarray}
As expected, when $\e0 = 1$, i.e., the detectors are completely correlated, the expressions for
false alarm and detection probability are consistent with those of a single detector. 
As $\e0$ decreases, the false alarm rate falls off more quickly and the detection
probability increases improving the performance of the network. In the other extreme limit,
when $\e0 = 0$, i.e., the detector noises are independent, the formulae reduce to those given by equations (\ref{ndetecfa}) and (\ref{detpr}).

%%%%%%%%%%%%%%%%%%%%%%%%%%%%%%%%%%%%%%%%%%%%%%%%%%%%%%%%%%%%%%%%%%%%%%%%%%%%%%%%%%%%%%%%%%%%%%%%%%%%

\subsection{Coincidence detection}
\label{facoin}

Coincidence detection is a simpler strategy where one detector does not `know' about the others - the detectors are treated as if in isolation. The data from the different detectors is processed separately;   separate thresholds are set and lists of crossings of the threshold or candidate events are recorded as separate lists. The lists are then compared for consistency in the parameters of the signal. Since a unique GW source is supposed to be responsible for the signal \cite{arch}, the estimated parameters of the signal in each detector must satisfy consistency requirements. Ideally, the times of arrival of the signal in the detectors should not exceed the light travel time between the detectors; in case of the spinless inspiral we consider here, in addition to the arrival times matching, the estimated mass parameters $\{\tau_0, \tau_3\}$ must be identical. However, because of the presence of noise in the detectors, the estimates of the parameters may differ from their true values and lie in a neighbourhood of the true values. Thus when deciding the consistency of the parameters of the signal one must account for the difference in estimates of the parameters in the different detectors. Specifically, in the case of two detectors, a detection will be inferred if corresponding to a crossing in the first detector, there is a candidate event in the second detector whose parameters lie in the 
neighbourhood of the parameters of the candidate event in the first detector. We call this neighbourhood which is a subset of the parameter space, a parameter window denoted by $\W$. For the specific situation we consider of identical, co-aligned and co-located detectors detecting an inspiral signal, a detection is inferred, if for a candidate event in detector 1  with estimated parameters $\{t_{c}^1, \tau_{0}^1, \tau_3^1 \}$, there is a candidate event in detector 2 with parameters $\{t_{c}^2, \tau_{0}^2, \tau_3^2 \}$ which lies in the parameter window $\W$, a box whose vertices are given by 
$\{t_c^1 \pm \Delta_{\W} t_c, \tau_0^1 \pm \Delta_{\W} \tau_0, \tau_3^1 \pm \Delta_{\W} \tau_3 \}$.
Recall that a candidate event in each detector is decided by the statistic in each detector crossing its  respective threshold. Note that we have taken the window to be symmetrical about the candidate event. This assumption is not unjustified on the basis of the simulations which we perform in the next section.     
\par 
The question arises as to how $\W$ is determined. One possible way is to use the Fisher information matrix in order to decide $\W$. The Fisher information matrix gives estimates of the errors in the parameters due to the noise. It also suggests that the errors scale inversely as the SNR. So the size of 
$\W$ depends on the SNR - the higher the SNR the smaller the size of $\W$, that is, the estimated values tend towards the true values of the parameters for large SNR. However, for low SNR $\sim 10$ or less, it has been shown that the Fisher information matrix is a poor estimator of the errors; it underestimates the errors by a large margin \cite{BSD} (this is shown for the inspiral waveform up to 1 PN order). Simulations show that the actual errors can be a factor 2 or even 3 larger than those predicted by the Fisher information matrix. We therefore use simulations (section IV) to empirically determine the parameter window $\W$. We find that the window size depends on the SNR and reduces with increasing SNR. Since the parameter space is sampled discretely with templates, 
$\W$ contains a finite number $\Nwin$ independent templates. From the foregoing discussion $\Nwin$ is a function of the SNR and decreases with increasing SNR. The quantity $\Nwin$ enters into the false alarm rate as we shall see below.    
 
\subsubsection{Uncorrelated noise}
\label{uncorr_coin}

Since each detector registers events on its own, the detection statistic for the first detector is just 
\break $\Lambda_1 = |C^1|^2$ (without any loss of generality, we may assume any one of the two detectors to be first). Given $\Nind$ number of independent templates which correspond to the three dimensional parameter space, the false alarm probability in detector 1 is given by:
\begin{equation}
P_{FA}^1 = \Nind \int_{\Ls}^\infty d\Lambda_1 p_0(\Lambda_1) = \Nind \hbox{exp} \left( - \frac{\Ls}{2} \right).
\end{equation}
Because the detectors are taken to be identical, the template banks in both are the same; the template bank depends only on the noise PSDs which are taken to be identical. Therefore we do not attach any subscript to $\Nind$.  We set the same threshold $\Ls$ in each detector. We then draw up lists of candidate events which cross the threshold $\Ls$ in each detector. We thus obtain two lists corresponding to each detector. We check for consistency between candidate events if a candidate event in detector 1 has at least one candidate event in detector 2 lying within the parameter window $\W$ centered around the parameters of the candidate event in detector 1. The size of $\W$ is determined by 
$\Ls$, as seen from the Fisher information matrix or otherwise through simulations. Therefore $\Nwin$ is a function of $\Ls$. Also by our choice of the parameters $\{\tau_0, \tau_3\}$, $\Nwin$ depends only weakly on these parameters. We ignore this dependence and consider $\Nwin$ only as a function of $\Ls$.  
Suppose that the number of templates within the parameter window is $N_{\rm win,t}(\Lambda^*)$. 
Then, we have $\Nwin(\Lambda^*)=\alpha(\Lambda^*) N_{\rm win,t}(\Lambda^*)$, where 
the factor $\alpha$ must be $0<\alpha(\Lambda^*)\leq 1$. 

The probability that at least one template among the $\Nwin(\Ls)$ templates registers a false alarm is \break $1 - \left(1-e^{-\Lambda^*/2}\right)^{\Nwin(\Lambda^*)}$. 
Thus the probability of false alarm for the two detector network is:
\begin{equation}
P_{FA} (\Ls) = \Nind e^{-\Lambda^*/2} \left[ 1 - \left(1-e^{-\Lambda^*/2}\right)^{\Nwin(\Lambda^*)} \right].
\label{pfa}
\end{equation}
This unwieldy expression for the false alarm probability for a high threshold $\Ls \gsim 50$, (which corresponds to an one detector SNR threshold of about 7) can be simplified to an approximate but simple expression: 
\begin{eqnarray}
P_{FA} (\Ls) &\simeq& \Nind \Nwin (\Lambda^*) e^{-\Lambda^*} \nonumber\\
&=&  \Nind \alpha(\Lambda^*) N_{\rm win,t}(\Lambda^*) e^{-\Lambda^*} \,.
\end{eqnarray}

The detection probability is given as follows: The false dismissal in each detector for a signal of amplitude $\A$ is given by:
\begin{equation}
P_{FD} = \h \int_0^{\Lambda^*} d \Lambda \hbox{exp} \left[-\frac{\Lambda + \A^2}{2} \right] 
I_0 (\A \sqrt{\Lambda}) \,.
\end{equation}
Then the detection probability of the network is just:
\begin{equation}
P_{DE} = (1 - P_{FD})^2 \, .
\end{equation}

\subsubsection{Correlated noise}
\label{2cocorr}

In order to compute the false alarm and false dismissal probabilities for this case,
we introduce a real correlation vector defined by, 
\be
\C^T = (c_0^1, c_{\pi/2}^1, c_0^2, c_{\pi/2}^2) \,.
\ee
The covariance matrix of $c^I_{0,\pi/2}, (I=1,2),$ are given by,  
\begin{equation}
\R \equiv \langle \C \C^T \rangle = 
\left( \begin{array}{cccc}
1 & 0 & \e0 & 0 \\
0 & 1 & 0 & \e0\\
\e0 & 0 & 1 & 0\\
0 & \e0 & 0 & 1
\end{array} \right) \,.
\label{cov}
\end{equation}
The procedure of the computation of the false alarm and the false dismissal
probabilities are as follows: 
we write down the pdf of $\C$, then integrate over the angular variables $\phi_I$ (as defined 
in Eq. (\ref{polar})) and obtain the pdfs 
of the quantities $\rho_I = |C^I| = \sqrt{\Lambda_I}$ in the absence and presence of the signal, namely, $p_0(\rho_1, \rho_2)$ and $p_1(\rho_1, \rho_2)$ respectively. From these pdfs the required false alarm and false dismissal probabilities are easily obtained on integration. 
\par
Before obtaining the pdf of $\C$, we require the determinant of $\R$ denoted by $\det \R$:
\be
\det \R = (1 - \e0^2)^2,  
\ee
and the inverse of $\R$:
\begin{equation}
R^{-1} = \frac{1}{\sqrt{\det \R}}
\left( \begin{array}{cccc}
1 & 0 & -\e0 & 0 \\
0 & 1 & 0 & -\e0\\
-\e0 & 0 & 1 & 0\\
0 & -\e0 & 0 & 1
\end{array} \right).
\end{equation}
\par
Under the null hypothesis $H_0$, when, no signal is present, the probability distribution of these
four quantities will be a multivariate normal distribution centered at the origin, i.e.,
\begin{equation}
p_0(\C) = \frac{1}{(2 \pi)^2 \sqrt{\det \R}} \exp \left[ -\frac{1}{2} \C^T \R^{-1} \C \right] \,.
\label{pdfC}
\end{equation}
We now go over to the `polar' variables $\rho_I$ and the angular variables $\phi_I$ by the relations:
\begin{equation}
c_0^I = \rho_I \cos \phi_I, \qquad c_{\pi/2}^I = \rho_I \sin \phi_I \qquad I=1,2.
\label{polar}
\end{equation}
$\rho_I$ is the registered SNR in the $I$-th detector and is compared with the same threshold $\rho^*$ in each detector. We marginalize the angular variables $\phi_I$ by integrating over them since it is the 
$\rho_I$ which are compared with the threshold in this type of search. We transform the pdf in Eq. (\ref{pdfC}) into the polar variables $(\rho_I, \phi_I)$ and integrate over $\phi_I$. The integral over $\phi_I$ leads again to the modified Bessel function $I_0$. After including the Jacobian factor, we obtain: 
\begin{equation}
p_0 (\rho_1, \rho_2) = \frac{\rho_1 \rho_2}{1-\e0^2} 
\exp \left[-\frac{\rho_1^2 +\rho_2^2}{2(1-\e0^2)} \right] 
I_0 \left(\frac{\e0 \rho_1 \rho_2}{1-\e0^2}  \right).
\end{equation}
Similar arguments that led to the expression in Eq. (\ref{pfa}) for the false alarm probability in the uncorrelated case need to be followed also in this case, except that one must be careful because of the correlated noise. The expressions therefore are more complex. We will require the Bayes theorem and conditional probabilities in order to compute the false alarm probability. We proceed as follows:
\par
We define the events $A$ and $B$ in the $(\rho_1, \rho_2)$ plane as follows:
\bea
\nonumber
A &\equiv& \{(\rho_1, \rho_2) | \rho_1 > \rho^*,~~ 0 < \rho_2 < \infty \}, \\
B &\equiv& \{(\rho_1, \rho_2) | 0 < \rho_1 < \infty,~~ \rho_2 < \rho^* \} \,. 
\eea 
In our computations we require the probabilities of the events $A \cap B$ and $A$, which are given in terms of the pdf $p_0 (\rho_1, \rho_2)$ as,
\bea
\nonumber
Q_1 &=& P (A \cap B) = \int_0^{\rho^*} d \rho_2 \int_{\rho^*}^{\infty} d \rho_1 p_0 (\rho_1, \rho_2) \,, \\
Q_2 &=& P (A) = \int_0^{\infty} d \rho_2 \int_{\rho^*}^{\infty} d \rho_1 p_0 (\rho_1, \rho_2) \,.
\eea
The event $A$ represents a false alarm occurring in detector 1 for one template, irrespective of detector 2 and $A \cap B$ represents the event that a false alarm occurs in detector 1 for one template and no false alarm occurs in detector 2. Then by Bayes theorem, the conditional probability that no false alarm occurs in detector 2 given that a false alarm occurs in detector 1 is :
\be
P (B/A) = \frac{P(A \cap B)}{P(A)} = \frac{Q_1}{Q_2} \,.
\ee
The probability that no false alarm occurs in $\W$ containing $\Nwin$ independent templates in detector 2 when a false alarm occurs in detector 1 is just $[P (B/A)]^{\Nwin}$. The probability that at least one false alarm occurs in detector 2 - say event $C$ - in the parameter window 
$\W$ given that a false alarm occurs in detector 1 is:
\be
P (C/A) = 1 - {\left( \frac{Q_1}{Q_2} \right)}^{\Nwin} \,.
\ee
The event that a false alarm occurs in detector 1 and at least one false alarm in $\W$ in detector 2 is 
$A \cap C$. The probability of this event is again given by Bayes theorem as:
\be
P(A \cap C) = P(A) P(C/A) = Q_2 \left[ 1 - {\left( \frac{Q_1}{Q_2} \right)}^{\Nwin} \right] \,.
\ee 
Recalling that there are $\Nind$ independent templates, the expression for the false alarm probability for coincident detection for the network becomes:
\begin{equation}
P_{FA} (\rho^*) = \Nind Q_2 \left[ 1 - {\left( \frac{Q_1}{Q_2} \right)}^{\Nwin} \right] \,.
\label{pfacc}
\end{equation}
The above expression goes over to Eq. (\ref{pfa}) of the uncorrelated case when $\e0$ approaches zero. Also in the limit of large thresholds $\rho^* >> 1$, we obtain a simple expression for the false alarm probability $P_{FA}$. Define the probability $Q_0$ by:
\be
Q_0 =  \int_{\rho^*}^{\infty} d \rho_2 \int_{\rho^*}^{\infty} d \rho_1 p_0 (\rho_1, \rho_2) \,.
\ee
Then $Q_2 = Q_1 + Q_0$ and as $\rho^* \longrightarrow \infty$, we see that $Q_0 \longrightarrow 0$. Then by making a simple expansion of $P_{FA}$ in Eq. (\ref{pfacc}) and keeping just the linear term in $Q_0$ we find that,
\be
P_{FA} (\rho^*) \simeq \Nind \Nwin Q_0 \,.
\ee  
This is the simple form we use in section IV to plot the ROC curves for which $\rho^* \gsim 7$.
\par
We now compute the detection probability. Under the alternative hypothesis $H_1$, when a signal of amplitude $\A$ is incident on the detectors, the mean of $\C$ is no more zero and we have a multivariate Gaussian distribution translated away from the origin. Explicitly, the means of its components are given by:
\begin{equation}
\langle c_0^I \rangle = \A \cos \phi_0, \qquad \langle c_{\pi/2}^I \rangle = \A \sin \phi_0 \,,
\end{equation}
where $\phi_0$ is the phase of the signal. Denoting the mean of $\C$ by $\C_1$ we obtain from the above,
\be
\langle \C^T \rangle = \C_1^T = \A ( \cos \phi_0, \sin \phi_0, \cos \phi_0, \sin \phi_0) \,.
\ee
 Then the distribution of $\C$ is given by,
\bea
\nonumber
p_1(\C) &=& \frac{1}{(2 \pi)^2 \sqrt{\det \R}} \\
 &\times& \exp \left[ -\frac{1}{2} (\C - \C_1)^T \R^{-1} (\C - \C_1) \right] \,,
\label{H1C}
\eea
The exponent in $p_1(\C)$ of Eq. (\ref{H1C}) can be written as $- Q /2(1 - \e0^2)$, 
where $Q$ is a quadratic in $\rho_I$ and is explicitly given in terms of the polar variables as: 
\begin{eqnarray}
\nonumber
Q & = & \rho_1^2 + \rho_2^2  + 2 \A^2 (1-\e0) - 2 \A (1-\e0)\\
\nonumber  & \times & [\rho_1 \cos(\phi_1-\phi_0) + \rho_2 \cos(\phi_2-\phi_0)]\\
  & - & 2 \e0 \rho_1 \rho_2 \cos (\phi_1 - \phi_2),
\end{eqnarray}

Integrating over the angular variables and multiplying by the Jacobian factor of $\rho_1 \rho_2$ we obtain implicitly the pdf in terms of $\rho_1, \rho_2$:
\begin{equation}
p_1 (\rho_1, \rho_2) = \rho_1 \rho_2 \int_{-\pi}^{\pi} \int_{-\pi}^{\pi} d\phi_1 d\phi_2 p_1(\C).
\label{H1rho}
\end{equation}
\par
We can obtain a useful approximation when the amplitude of the incident wave is large, i.e. \break $\A >>1$. We evaluate the above integral using \emph{stationary phase approximation}. Since the result is independent of $\phi_0$, the phase of the signal, we may without loss of generality put it equal to zero. Then the integrand contributes only in the neighbourhood of $\phi_1 = \phi_2 = 0$. Expanding $Q$ up to second order in $\phi_I$ we obtain:
\begin{eqnarray}
\nonumber
Q & \simeq & \sum_{I=1}^2 \left[\rho_I^2 - 2 \A (1-\e0) \rho_I (1- \frac{\phi_I^2}{2})\right]\\
\nonumber & + & 2 \A^2 (1-\e0) \\
          &-& 2 \e0 \rho_1 \rho_2  \left[1-\frac{(\phi_1 - \phi_2)^2}{2}\right].
\end{eqnarray}
%Leaving out the terms dependent on $\phi_I$, which will contribute only to an overall
%normalization constant, the probability distribution function in Eq. (\ref{H1rho}) reduces to the familiar Gaussian one centered at the amplitude in both detector dimensions. 
Since the contribution to the integral over $\phi_I$ comes from the region mainly near $\phi_I=0$, 
we can change the integration region from $[-\pi,\pi]$ to $[-\infty,+\infty]$. 
We thus obtain: 
\begin{widetext}
\begin{equation}
p_1 (\rho_1, \rho_2) \simeq \frac{\sqrt{\rho_1\rho_2}}{2 \pi (1-\e0)\A} 
\left[1+\frac{\e0(\rho_1+\rho_2)}{(1-\e0) \A}\right]^{-1/2} 
\exp \left[-\frac{(\rho_1-\A)^2 + (\rho_2-\A)^2 -2\e0(\rho_1-\A)(\rho_2-\A)}{2(1-\e0^2)}\right] \,.
\end{equation}
\end{widetext}
This is similar to the familiar Gaussian distribution centered at the amplitude in both detector dimensions. 
Because of the correlation, the Gaussian is tilted at an angle of $45^\circ$ with standard deviations 
$\sqrt{1 \pm \e0}$ along the two eigendirections. 
\par
The detection probability is then given by integrating from $\rho^*$ to $\infty$ along both detector dimensions. Thus, 
\begin{equation}
P_{DE} = \int_{\rho^*}^\infty d\rho_1 \int_{\rho^*}^\infty d\rho_2 p_1 (\rho_1, \rho_2).
\end{equation}

%%%%%%%%%%%%%%%%%%%%%%%%%%%%%%%%%%%%%%%%%%%%%%%%%%%%%%%%%%%%%%%%%%%%%%%%%%%%%%%%%%

\section{Numerical results from simulations}
\label{sim}

In the previous section we obtained useful expressions for the false alarm and detection probabilities in  the coherent and coincidence detection in the uncorrelated and correlated noise cases. These probabilities are required in the plotting of ROC curves in each of the four cases. However, the false alarm probabilities in each case Eqs.(\ref{ndetecfa}), (\ref{pfacohc}), (\ref{pfa}), (\ref{pfacc}) depend on two undetermined quantities namely,  $\Nwin$ the size of the parameter window $\W$ in the coincidence cases and $\Nind$, the number of independent templates in all cases. These quantities are difficult to determine analytically and must be obtained from simulations. Therefore the first goal of this section is to estimate these quantities and in doing so obtain the template density in $\{\tau_0, \tau_3 \}$ for the 3 PN inspiral waveform. The template density is required to compute the number of templates $\Nwin$ in $\W$. The final goal of this section and also the paper is to plot the ROC curves for each of these cases and compare them for performance metrics. 

\subsection{The template bank for the 3 PN waveform}
\label{temp}
At high minimal matches $MM$ close to unity, the template bank is obtained conveniently from the metric on the intrinsic parameter space $\{\tau_0, \tau_3 \}$ - the quadratic approximation near the peak of the ambiguity function is adequate. Recent work in \cite{arun} deals with 3.5 PN inspiral waveform and implicitly computes the Fisher information matrix which is in fact the metric on the full parameter space which includes other kinematical parameters on which the waveform depends. The idea of a metric was introduced in \cite{BSD} and its computation on the space of the two masses \cite{owen} - the intrinsic parameter space - has been amply dealt with in the literature. Here we do not go into the details of its computation but directly state the results. 
\par
The metric depends on the noise PSD of the detectors. Since we take identical detectors the noise PSD is the same. We take this to be that of initial LIGO as given by the analytic approximation in \cite {DIS}. 
Our final results are not very sensitive to the exact PSD and therefore it is adequate to employ the analytic approximation which is convenient to use in our simulations. The expression is: 
\begin{widetext}
\begin{eqnarray}
\nonumber
S_h(f) & = & S_0 \left[ (4.49x)^{-56} +0.16x^{-4.52} + 0.52 + 0.32x^2 \right] \qquad  f \geq f_s \\
& = & \infty \qquad \hbox{otherwise},
\end{eqnarray}
\end{widetext}
where, $x = f/f_k$, $f_k = 150$Hz, $f_s = 40$Hz,  and $S_0 = 9 \times 10^{-46}$/Hz.
\par
The metric on the parameter space $\{\tau_0, \tau_3\}$ is denoted by $g_{\alpha \beta}$ 
where the indices $\alpha, \beta$ refer to the coordinates $ \lambda^{\alpha} \equiv \{\tau_0, \tau_3 \}$. The significance of the metric is that the "squared distance" between adjacent normalized templates separated by the coordinate vector $\Dl^{\alpha}$ is given by $g_{\alpha \beta} \Dl^{\alpha} \Dl^{\beta}$. 

\begin{figure}[h!]
\vskip 0.6cm
\centering
\includegraphics[width=0.45\textwidth]{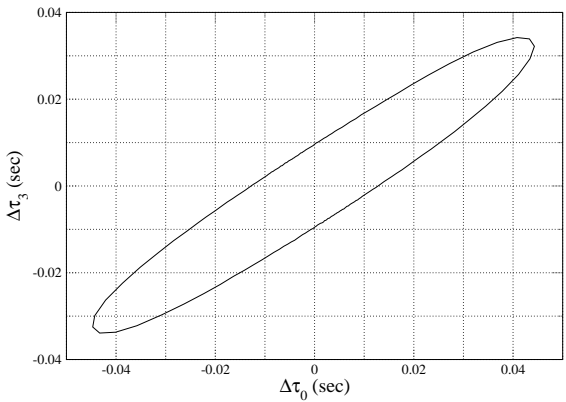} 
\caption{Plot of the contour of the ambiguity function at $MM = 0.97$. The fiducial frequency which scales both the axes is $f_a = 40$ Hz.}
\label{ambgty}
\end{figure}

This squared distance is determined by the fractional mismatch that one is prepared to tolerate i.e.:
\be
g_{\alpha \beta} \Dl^{\alpha} \Dl^{\beta} = 1 - MM \,.
\label{metric}
\ee
where MM is called the minimal match.
The Fig.\ref{ambgty} shows the contour of the templates which are at a fractional match of 0.97 from the central template ($\Delta \tau_0 = \Delta \tau_3 = 0$) for the 3 PN inspiral waveform. This is in fact the  contour of the ambiguity function at the level $MM = 0.97$.
 In general the metric would depend on the location $(\tau_0, \tau_3)$ in the parameter space; however, the specific parameters $\{\tau_0, \tau_3 \}$ have been chosen so that the metric is almost constant over the full parameter space. That this is possible is a consequence of the fact that the parameter space is almost flat. 
\par
We may construct a template bank with the templates placed parallel to $\tau_0$ and $\tau_3$ axes. 
%This is done for example in the LAL routines \cite{lal}. 
For $MM = 0.97$, the distance $\Dt_0$ and $\Dt_3$ between the adjacent templates in each direction $\tau_0$ and $\tau_3$ respectively is    
$\Dt_0 \sim 25$ms and  $\Dt_3 \sim 19$ms. For the purposes of our simulations it is sufficient to select a `small' rectangle compared with the full parameter space which usually covers a region corresponding to $M_l \le m_1, m_2 \le M_u$, where $M_l \lsim 1 M_\odot$ and $M_u \sim 30 M_\odot$. By doing so, we have the important benefit of saving enormously on the computational costs, without sacrificing on the results as the metric is nearly flat. We thus assume an uniform distribution of templates in the selected rectangle, which has a corner on the point $(\tau_0, \tau_3)$ corresponding to each component mass equal to $1.4 M_{\odot}$. We choose the rectangle large enough to accommodate 625 templates.

\subsection{Estimating $\Nind$}
\label{deltat}

Since the data is sampled at $\Delta^{-1} \gsim 2048$ Hz, it produces time samples in the statistic with sampling interval $\Delta \lsim 0.5$ ms. Thus the statistic is very finely sampled in $t_c$ which gives rise to strong correlations between adjacent samples. Although this is true for the full parameter space the correlations are particularly strong in the parameter $t_c$. These correlations tend to give us erroneous estimates of the false alarm probability if we consider all the samples as independent random variables. We need to factor in this effect of correlated output to estimate the number of independent templates $\Nind$.  A single detector analysis is sufficient.
\par
\begin{figure}[h!]
\vskip 0.6cm
\centering
\includegraphics[width=0.45\textwidth]{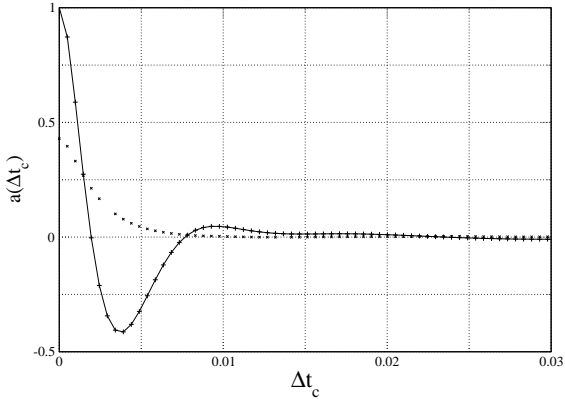} 
\caption{Plots of the auto-correlation functions $c(\Delta t_c)$ (dotted curve) and $c_0(\Delta t_c)$ (solid curve). For reference, the autocorrelation curve for one of the quadrature components $c_0$ is also plotted. This curve is obtained both analytically in \cite{DS94} and here numerically through simulations. The curve for $c$ is obtained purely numerically via simulations.}
\label{autcr}
\end{figure}

It is known that the filtered output passed through a band-pass filter is correlated. It was shown in 
\cite{DS94} that the chirp filter (taken to be Newtonian here for simplicity) acts effectively as a band-pass filter with the lower limit frequency as the seismic cut-off which acts like a `wall' and the upper frequency determined by the signal power fall off which is $\sim f^{-7/3}$ for the chirp waveform. Even if the raw noise is white, the chirp filter output for successive $t_c$ is correlated, because the band-pass filter effectively reduces the number of degrees of freedom in the original raw data. In the figure below we plot the auto-correlation function of the filtered output $c(t_c)$ as a function of $\Delta t_c$, namely, $a(\Delta t_c) = \langle c(t_c) c(t_c + \Delta t_c) \rangle$ which is maximized over the initial phase but with the same mass parameters. The $\Delta t_c$ at which the autocorrelation approaches zero is the required \emph{decorrelation time}. From the Fig. (\ref{autcr}) it appears that the decorrelation time $\sim 15$ms when the autocorrelation function falls to few percent of its maximum value.  
\par
\begin{figure}[h!]
\includegraphics[width=.35\textwidth, angle=270]{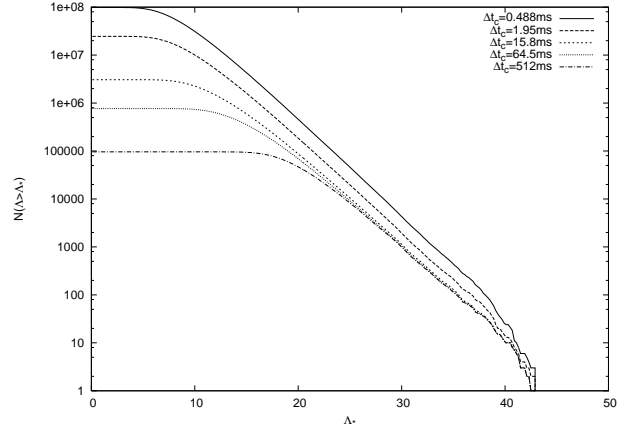} 
\caption{The number of false alarms $N_{FA}$ as a function of the threshold $\Lambda^*$ for various fixed  values of $\Delta t_c$.}
\label{fa}
\end{figure}

More directly, relevant to the false alarm probability computation, we evaluate the decorrelation time of the filtered output by the following procedure. We take data trains 32 sec long which are sampled at 2048 Hz giving 65536 points in a data train. We take 1500 realizations of the noise and pass the data through 
500 templates in the rectangle. We divide the matched filter output into data chunks of various lengths starting from the sampling interval of 0.488 ms up to 512 ms increasing each time the length of the data chunk. The size of the chunk is the trial decorrelation time. For each fixed length of the chunk (or trial decorrelation time) we maximize the statistic 
$\Lambda_1 = |C^1|^2$ over each chunk and the 500 templates, and compare it with a threshold $\Lambda^*$. We count the number of crossings $N_{FA}$ of the statistic $\Lambda_1 > \Lambda^*$ - the number of false alarms. We take the range $0 < \Lambda^* < 50$ and obtain $N_{FA}$ as a function of $\Lambda^*$. The number of false alarms can range up to $32 \times 2048 \times 1500 \sim 10^8$ 
for the case $\Delta t_c=0.488$ms. 
We then plot the curve $N_{FA}$ as a function of $\Lambda^*$ for a fixed data chunk of length $\Delta t_c$. We now increase the length of the chunk $\Delta t_c$ from the sampling rate of $\sim 0.488$ ms to 512 ms successively and plot the curves for various chunk lengths (trial decorrelation times) $\Delta t_c$. We find that for the decorrelation times $\Delta t_c \geq \Delta t_{c \infty} \sim 15.8$ ms the curves start to coalesce into a single curve in the large $\Lambda^*$ regime. This signifies that the samples spaced 
$\Delta t_c \gsim \Delta t_{c \infty}$ apart are essentially decorrelated and will produce satisfactory results. This procedure is depicted in Fig. \ref{fa} by plotting some of the curves with $\ln N_{FA}$ versus $\Lambda^*$ for various fixed values of $\Delta t_c$. 

However, in order to obtain good statistical results, it is desirable to choose the maximum number of data chunks. We satisfy both these criteria by choosing the data chunk length to be the least, but at the same time ensuring decorrelation in samples, namely, we choose the chunk length to be $\Delta t_{c \infty}$.
From the simulations we have also verified that the autocorrelation function $a(\Delta t_c)$ falls off to a value of about $3\%$ of its maximum value at $\Delta t_c = 15.8$ ms (the figure (\ref{autcr}) also suggests this fact). 
\par
A similar procedure is followed for evaluation of the decorrelation length in the parameters 
$\tau_0$ and $\tau_3$. We find that the samples spaced  
$\Delta\tau_0\geq 220$ms apart for $\tau_0$ and $\Delta\tau_3\geq 80$ms 
for $\tau_3$, are decorrelated and will produce satisfactory results.
The template space for our simulation has the length $600$ms for $\tau_0$
and $450$ms for $\tau_3$. We then take the maximum among the samples
with different $\tau_0, \tau_3$ parameters. 
\par
The number of false alarms  at sufficiently high values of the threshold are given by:
\begin{equation}
N_{FA}(\Lambda > \Lambda^*)) = N_{\rm sim} \Nind \exp \left(-\frac{\Lambda^*}{2}\right),
\label{intrcpt}
\end{equation}
where, $N_{\rm sim} $ is the number of simulations we perform. We plot $\ln N_{FA}$ versus 
$\Lambda^*$ in Fig. \ref{fa2} below.
\begin{figure}[h!]
\centering
\includegraphics[width=.5\textwidth,angle=0]{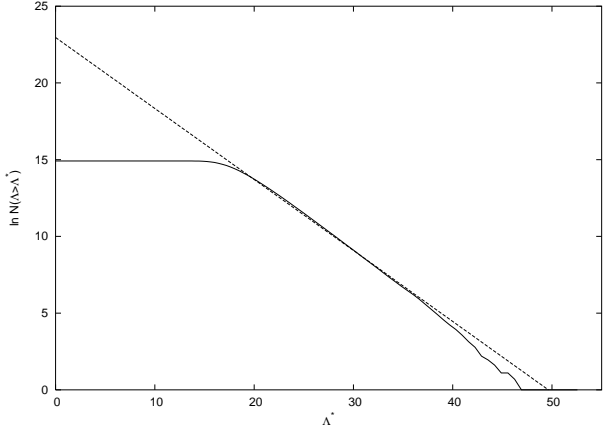} 
\caption{The natural log of the number of false alarms is plotted versus the threshold $\Lambda^*$. In the large threshold regime, a straight line is fitted to this curve and extended to obtain intercepts on both axes. The intercept on the vertical axis is $\sim 22.96$.}
\label{fa2}
\end{figure}
%\vskip 1cm
In this plot, we perform $N_{\rm sim}=1500$ simulations. The data length is 32 seconds with a sampling rate of 2048Hz. A rectangle with  $50 \times 10 = 500$ templates was taken in the $(\tau_0, \tau_3)$ plane.   
%The length in the mass parameter region is 0.6 seconds for $\tau_0$ direction and 0.46 seconds 
%for $\tau_3$.

We observe that the graph of $\ln N_{FA}$ versus $\Lambda^*$ is a straight line when $\Lambda^*$ is sufficiently large. We use the least square fit in order to fit a straight line in the regime of large 
$\Lambda^*$, namely, in the range $17 \le \Lambda^* \le 37$. The equation of this straight line is:
\be
\ln N_{FA} = - 0.46 \Lambda^* + 22.96.
\ee  
Then $\Nind$ can be determined by extending the straight line to obtain intercepts on either axis. We find that the intercept on the $\ln N_{FA}$-axis is at $\ln N_{FA} = 22.96$. 
From Eq. (\ref{intrcpt}), we  conclude that $\Nind \sim e^{22.96}/1500 \sim 6.24  \times 10^6$ which is about $19\%$ of the total number of templates $32 \times 2048 \times 500 = 3.28\times10^7$ spanning the three dimensional parameter space $\{t_c, \tau_0, \tau_3\}$. 

\subsection{Determining the parameter window $\W$}
\label{windowfn}

In this section we discuss the size of the parameter-window which is required to decide coincident detection. Let a signal be detected individually by the two detectors. Even though the detectors are assumed to be identical with respect to the noise PSD, location and orientation, the different noise realizations in the two detectors will in general produce different estimates of the true parameters of the signal
\footnote{The calibration error in each detector may add to the difference in the estimates of the parameters. However, at SNR$\lesssim 10$, this effect can be ignored in the current detectors such as TAMA300. }. 
If the SNR is sufficiently high $\sim 10$ then presumably the detected parameters will not differ too much from the true parameters of the signal, and thus also not too much from each other. We therefore need to decide on the size of a box or a window in the parameter space, in order to say whether it is the same signal from an unique astrophysical source detected in each detector. Taking into account the reasons mentioned in section III B, we determine the size of $\W$ by performing simulations. 
\par
\begin{figure}[h!]
\includegraphics[width=.45\textwidth]{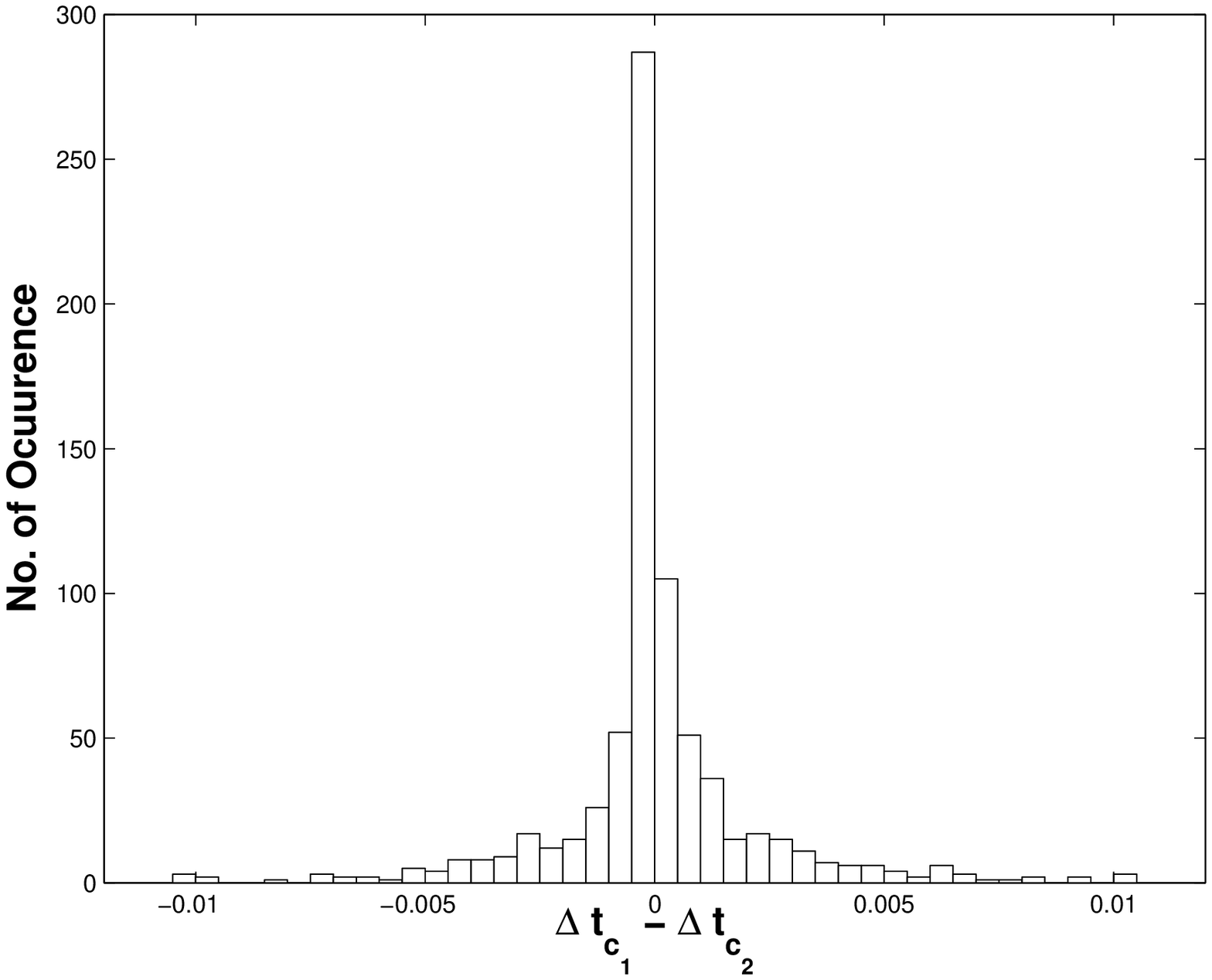}  
\mbox{(a)}
\includegraphics[width=.45\textwidth]{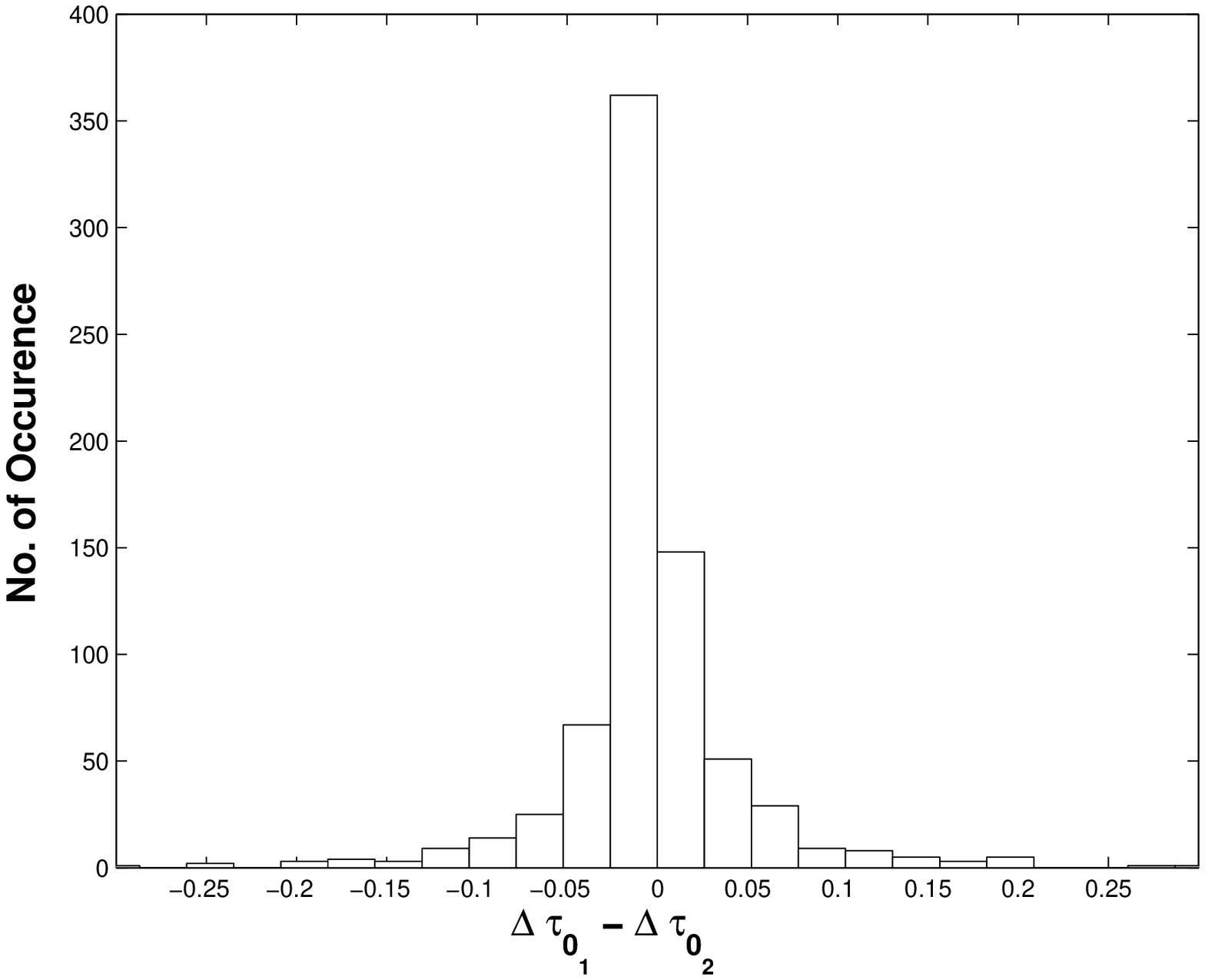} 
\mbox{(b)}
\includegraphics[width=.45\textwidth]{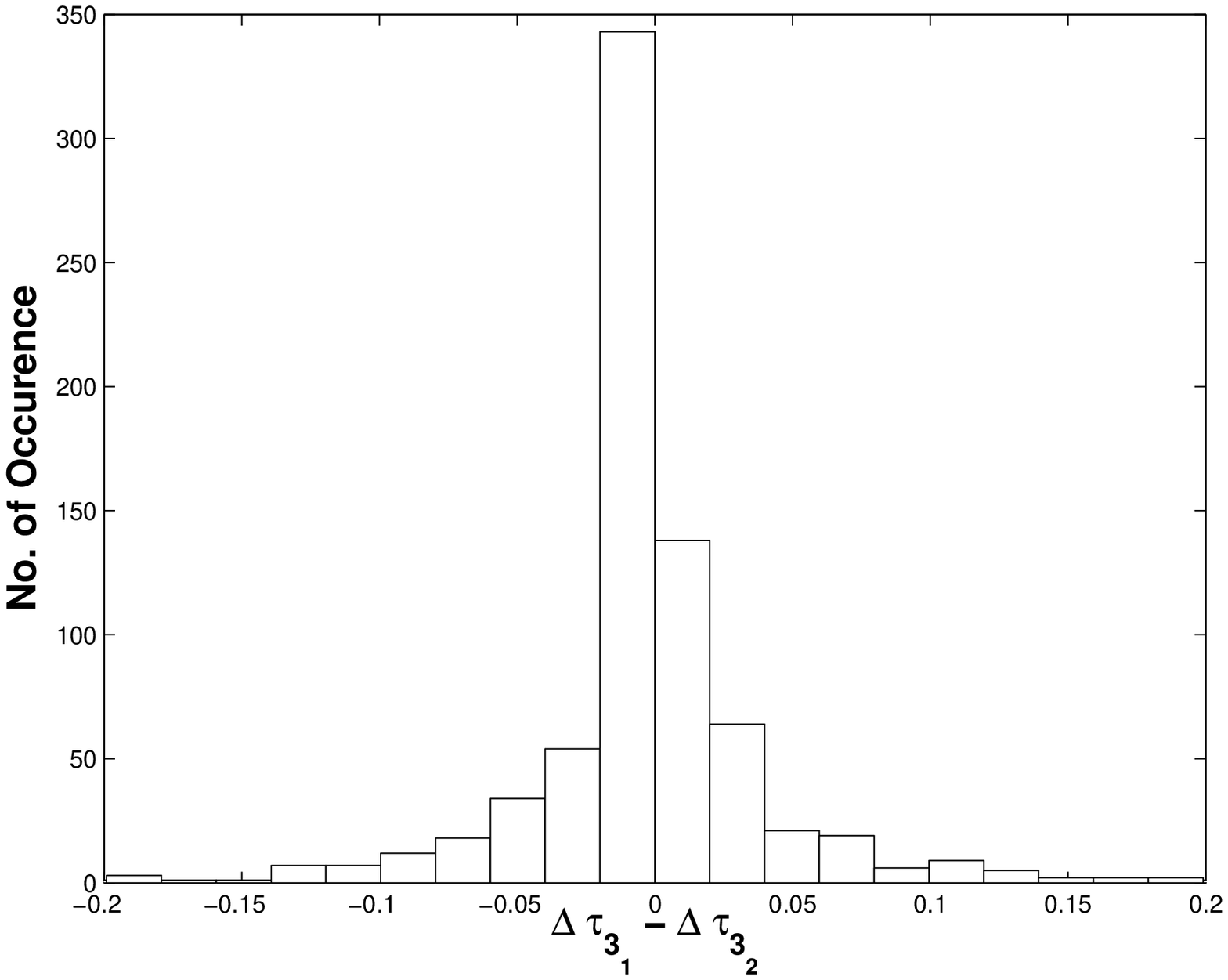} 
\mbox{(c)}
\caption{Distribution of the difference of the detected parameters, 
(a) $t_c$, (b) $\tau_0$ and (c) $\tau_3$, for an injected signal of SNR=10.}
\label{dist1}
\end{figure}

As seen from the arguments pertaining to the Fisher information matrix or from the simulation results, the size of $\W$ depends on the SNR of the signal - the higher the SNR, the smaller the size of $\W$ - the error in the estimates reduces. For a signal injected with typical parameters (lying somewhere near the center of the parameter space) with typical SNR = 10, the Fig. \ref{dist1} shows the distribution of the difference $\{\Delta t_c, \Delta \tau_0, \Delta \tau_3\}$ of the parameters in two detectors.

750 simulations have been performed for two detectors and the results are plotted as histograms. 
The vertical axis shows the number of events detected in each bin for each of the parameters in the three figures. The distributions taper off as we go away from the true signal parameters.  We fix the size of 
$\W$ by requiring that at least $99\%$ of the observed events in each parameter fall in the intervals 
$|\Delta t_c| \leq \Delta_{\W} t_c$, $|\Delta \tau_0| \leq \Delta_{\W} \tau_0$ and $|\Delta \tau_3| \leq \Delta_{\W} \tau_3$. This requirement fixes the window size. The percentage $99\%$ is subjective but seems to us to be reasonable. We therefore have:  
\begin{widetext}
\be
\W = \{(\Delta t_c, \Delta \tau_0, \Delta \tau_3) | |\Delta t_c| \leq \Delta_{\W} t_c,~ |\Delta \tau_0| \leq \Delta_{\W} \tau_0, ~|\Delta \tau_3| \leq \Delta_{\W} \tau_3 \}
\ee
\end{widetext}
The above results were obtained for a typical SNR of 10. Because of noise, even if the signal has true SNR of 10, the observed SNR may differ. For a signal with true SNR = 10, the observed SNRs are found to vary anywhere from roughly 7 to 14. This is shown in  Fig. (\ref{dist2}) below. 

\begin{figure}[h!]
\centering
\includegraphics[width=.45\textwidth]{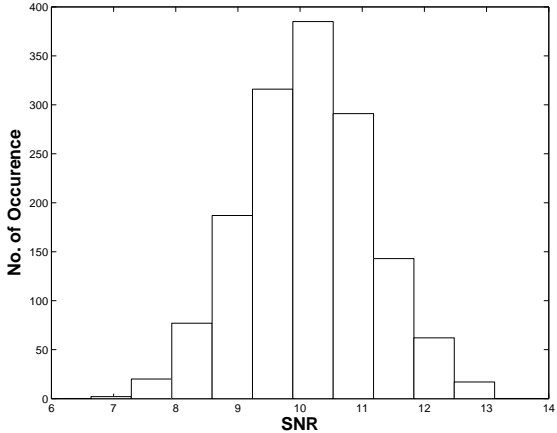} 
\caption{Distribution of SNR for an injected signal of SNR=10.}
\label{dist2}
\end{figure}

However, $\W$ depends on the SNR. So for the three parameters we determine $\W$ for various SNRs ranging from 7 to 14 in steps of unity. The window-sizes were taken such that at least $99\%$ of the observed values in each dimension fall within the window. This procedure detects the signal on an average 
$(0.99)^3 \sim 0.97$ of the time.
\par
The following table (\ref{win}) shows the results of our simulations. 750 signals of a specific SNR, say 7, are injected in the simulated noise of two detectors and then the number of events within a given bin size are counted and this number is normalized to unity by dividing by 750. This simulation obtains for us a window-size in each parameter (three in all for the three parameters) for SNR = 7. The procedure is repeated in addition for SNRs ranging from 8 to 14 in steps of unity. From the table we may infer the window size in each parameter for the various SNRs. The window size is tabulated in units of the bin size, namely, the distance between templates, for each of the parameters. We err on the safer side by choosing the bin size to be the smallest integral value larger than the corresponding fractional bin size for which the cumulative frequency crosses the 0.99 mark. 

%\begin{figure}[h!]
%\vskip 0.5cm
%\centering
%\includegraphics[width=0.42\textwidth]{tc_window.eps}
%\hbox{(a)} 
%\vskip 0.8cm
%\includegraphics[width=0.42\textwidth]{tau0win.eps} 
%\hbox{(b)}
%\vskip 0.8cm
%\includegraphics[width=0.42\textwidth]{tau3win.eps} 
%\hbox{(c)}
%\vskip 0.5cm
%\caption{Distribution of normalised cumulative frequencies for determining window size in the parameters  (a) $t_c$, (b) $\tau_0$ and (c) $\tau_3$ for SNRs ranging from 7 to 14 in steps of unity. `Upper' curves correspond to higher SNR. A horizontal dotted line in each figure depicts the $99\%$ level.}
%\label{win}
		%\end{figure}

%\begin{widetext}
\begin{table}
\begin{tabular}{|c|c|c|c|}
\hline
SNR & $\Delta_{\W} t_c$ (in units& $\Delta_{\W} \tau_0$(in units & $\Delta_{\W} \tau_3$(in units\\
& of 1/2048 s) & of 25ms) & of 19ms) \\
\hline
14 & 7 & 3 & 2\\
13 & 7 & 3 & 3\\
12 & 8 & 5 & 4\\
11 & 8 & 7& 5\\
10 & 10 & 8 & 5\\
9 & 12 & 10 & 7\\
8 & 14 & 12 & 8\\
7 & 17 & 15 & 10\\
\hline
\end{tabular}
\caption{Window size in the parameters $t_c$, $\tau_0$ and $\tau_3$ for SNRs ranging from 7 to 14 in steps of unity. Window
size corresponds to 99\% detection.}
\label{win}
\end{table}

%By counting the number of templates within $\W$ for a given SNR determines the quantity $\Nwin$. Thus now 
%we have all the relevant quantities for plotting the ROC curves. We do so in the next subsection. 

The window size $\W$ determines the number of template $N_{\rm win,t}$ within the windows. 
In order to plot the ROC curves, we need to estimate $\Nwin$, or equivalently $\alpha$. 
However we find that it is not easy to determine $\alpha$ as we did to determine $\Nind$. 
We thus need the help of the full numerical simulation  to determine $\alpha$, which is discussed in
the next subsection.

\subsection{The ROC curves}
\label{roccurves}

As mentioned before, the ROC curves fairly compare the coherent and coincidence strategies of detecting signals. At the same false alarm we can compare the detection efficiency (probability) by plotting these curves. We consider several situations here. The simplest of these involve injecting signals in a small rectangle in the $(\tau_0, \tau_3)$ space in the neighbourhood around the point corresponding to individual masses of $1.4 M_{\odot}$. We plot the ROC curves for the uncorrelated case $\e0 = 0$ and the correlated cases $\e0 = 0.2, 0.3$. We then extrapolate our results to the full parameter space by making certain assumptions which we state later and plot the ROC curves. Finally, we consider a uniform distribution of sources, uniform in spatial density and orientation with individual masses distributed close to $1.4 M_{\odot}$ and compare the performance of the two strategies. 
\par
The signal is injected in a rectangle $\Rs$ of the $(\tau_0, \tau_3)$ plane with its center at 
$\tau_0 = 25.1$ sec , $\tau_3 = 1.0743$ sec which lies well inside the parameter space, and away from its boundaries. The signal parameters are drawn randomly from a uniform distribution over the section of the parameter space consisting of the rectangle $\Rs$ in the $(\tau_0, \tau_3)$ space, $t_c$ and initial phase. The SNR is fixed at 10. In order to detect a signal in $\Rs$, the detection templates also span a rectangle $\Rt$, larger than $\Rs$ to take into account effects of noise - a signal lying in $\Rs$ could be detected by a template outside it, especially if the signal lies close to its boundary. $\Rt$ is chosen sufficiently large so that the width of the margin $\Rt - \Rs$ is $\sim \Delta_{\W} \tau_0$ and $ \sim \Delta_{\W} \tau_3$ in the relevant dimensions. 
\par
Specifically, we take $\Rs = (24.9375 \rightarrow 25.2625)~ {\rm sec} \times (0.9603 \rightarrow 1.2073)$ sec and $\Rt = (24.8 \rightarrow 25.4)~ {\rm sec} \times (0.8463 \rightarrow 1.3023)$ sec.
\par
\begin{figure}[h!]
\vskip 0.5cm
\centering
\includegraphics[width=0.43\textwidth]{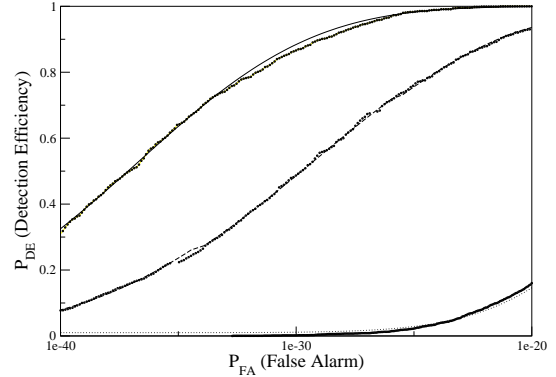} 
\caption{The ROC curves for single detector, uncorrelated coincident and coherent analysis for 
an injected signal of SNR=10.
The solid line, dashed line and the dotted line correspond to the theoretical ROC curves 
for coherent, coincident and single detectors 
derived in Section III, with $\Nind$ determined in Section IV.B. We also take $\alpha=1$ 
irrespective of the threshold. } 
\label{roc}
\end{figure}
\begin{figure}
\centering
\includegraphics[width=0.45\textwidth]{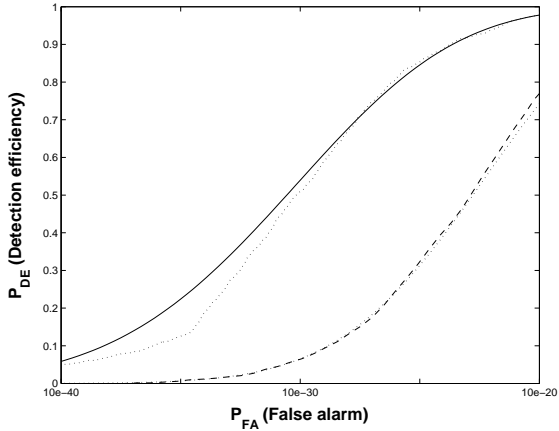} 
\hbox{(a)}
\includegraphics[width=0.45\textwidth]{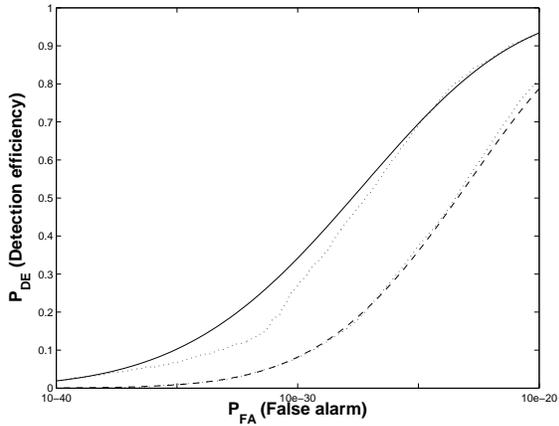} 
\hbox{(b)}
\caption{The ROC curves for coherent and coincidence detection for two correlated detectors with 
$\e0$ values (a) 0.2 and (b) 0.3 for an injected signal of SNR=10.} 
\label{roccorr}
\end{figure}

The resulting figures are plotted for $\e0 = 0$, 0.2 and 0.3 (figures \ref{roc}, \ref{roccorr}(a) and \ref{roccorr}(b), respectively). The trend of declining performance for increasing
values of $\e0$ is evident. For the sake of comparison, we have also
plotted the ROC curve of a single detector. However,
the performance of coherent detection is better than that of the coincident detection for all thresholds. We get a factor of roughly 2 to 4 times improvement in the 
detection efficiency over the coincident detector for a given rate
of false alarm. It justifies, why despite the heavier computational burden \cite{arch}, the coherent detection strategy is preferable. It allows one to
push the false alarm rate down for a given detection efficiency, increasing the confidence
of detection. In Figs.\ref{roc} and \ref{roccorr}, theoretical ROC curves derived in Section III are also plotted. We use the value of $\Nind$ determined 
in Section IV.B. On the other hand, we set $\alpha=1$ (i.e., $\Nwin=N_{\rm win,t}$) for the value of window parameters. We find that with this choice of the value $\Nind$ and $\alpha$, we can obtain a fairly good fit of theoretical curves to the simulation curves. 
As discussed in Section IV.C, the size of parameter windows and the decorrelation length suggests  that the number of independent templates in the parameter window is nearly 1 (i.e., $\alpha\simeq 1/N_{\rm win,t}$). Although a small change in $\alpha$ does not change the ROC curve very much, 
if we set $\alpha\simeq 1/N_{\rm win,t}$, the theoretical ROC curve deviates from the simulation curve significantly.  
This suggests that although the templates within a parameter window in Table I are spaced 
within the decorrelation length, the correlations between samples are not very large. 
We thus need to take $\alpha$ which is slightly smaller than unity, $\alpha\lesssim 1$. 

Although, it is not very easy to explain the reason for this choice of $\alpha$ for coincident detection, since the theoretical ROC curves and simulated ROC curves agree very well, 
we can use the theoretical curves for the evaluation of the performance of 
the coherent and coincident search of a much larger data set. 
We consider a one year data train and search the whole parameter space
spanned by masses 1$M_\odot$ - 40$M_\odot$. We assume that the sampling rate 
of the detector is 2048Hz. Then the number of time templates would be
\begin{equation}
N_{\rm t} = \frac {\hbox{length of interval}}{\hbox{sampling interval}} 
\sim \frac{3\times10^7 \hbox{sec}}{4.9\times10^{-4}\hbox{sec}} \simeq 6.1\times 10^{10}.
\end{equation}

\begin{figure}[h!]
\vskip 0.5cm
\centering
\includegraphics[width=.42\textwidth]{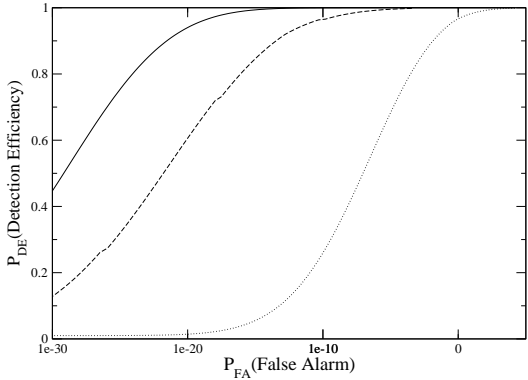} 
\caption{ROC curves for single(dotted line) coherent(solid line) and coincident(dashed line) detectors (uncorrelated) for a 1 year data train sampled at 2048Hz and searched over $1M_{\odot}-40M_{\odot}$, for an injected signal of SNR=10.}
\label{1yr}
\end{figure}

In order to evaluate the number of mass templates, 
we assume the metric is constant in the parameter space, for simplicity. 
We follow the analysis by Owen \cite{Owen}. 
For the LIGO I noise spectrum, 3PN phasing formula and $MM=0.97$, 
we obtain $N_{\rm m} \sim 1.2 \times 10^4$. 
Thus, the total number of templates is
\begin{equation}
N_{\rm tot} = N_{\rm t} N_{\rm m} \sim 7.3 \times 10^{14}.
\end{equation}

%Though actual simulation for such a huge number of templates will require enormous computing resources, we can still plot the ROC curves. 
%The only unknown quantity is $N_{ind}$, the number of 
%statistically independent templates, which appears in the expressions of false alarm rate and detection
%efficiency. 
In section IV.B, we have determined the value of this fraction of independent templates to the number of total templates to be 0.19. As only the parameter space has enlarged and the metric is nearly flat (which implies the correlations between templates depend essentially on the difference between parameters and not on the locations of the templates), the fraction of independent templates to the total number of templates
should not be very different from this value - we assume this fraction to be the same, namely, 0.19. 
We also assume $\alpha=1$ for the coincident detection. 
Assuming $\Nind$ and $\alpha$ in this manner, and SNR = 10, we plot the ROC curves in Fig. \ref{1yr}. We find that, in fact, the net effect is to only translate the ROC curves along the $P_{FA}$ axis relative to the ones plotted in the previous cases. 
\par
\begin{figure}[h!]
\vskip 0.5cm
\centering
\includegraphics[width=.42\textwidth]{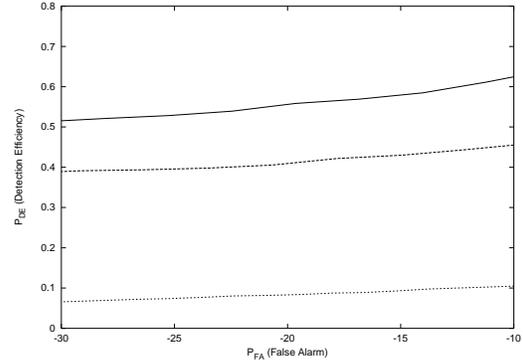} 
\caption{ROC curves for single(dotted line) coherent(solid line) and coincident(dashed line) detectors (uncorrelated) for a 1 year data train. The sources are distributed uniformly in distance between 
0.4 Mpc and 26 Mpc and also in orientation. }
\label{ROCdistributedsource}
\end{figure}

Finally, we consider the case of inspiral sources distributed uniformly in space and uniformly in orientation - the direction of the orbital angular momentum of the binary is distributed uniformly over the sphere.  We assume a uniform distribution within 0.4 Mpc and 26 Mpc. The lower limit on the distance ensures that our galaxy is excluded and the upper limit on the distance comes from the maximum range of LIGO I detectors for SNR of 7.  The direction of the orbital angular momentum is determined randomly. 
We consider the one year data train, sampled at 2048Hz and the mass parameter region, $1-40M_{\odot}$.  
The ROC curves are given by Fig.\ref{ROCdistributedsource}. We find that the coherent detection is better than the coincident detection by about 25-40\%. 

%%%%%%%%%%%%%%%%%%%%%%%%%%%%%%%%%%%%%%%%%%%%%%%%%%%%%%%%%%%%%%%%%%%%%%%%%%%%%%%%%%

\section{Conclusion}
\label{remark}

We have compared two possible strategies of multi-detector detection, namely, the coincidence and the coherent. We have considered the inspiral waveform for this purpose and by plotting the ROC curves for the two strategies we obtain a fair assessment of their performance. We have considered several cases from the simple to implement, to the more astrophysically relevant one. The bottom line here is that the coherent strategy of detection is superior to the coincidence strategy. The ROC curves display this fact quantitatively for all the cases considered. The coherent detection strategy uses the likelihood function which inherently uses the information of phase coherence to decide on detection. This information is encoded in the pdf of the $H_1$ hypothesis, that a signal is present in the data. The pdf explicitly contains the signal from each detector in the network with consistent phase. Thus the coherent strategy inherently accommodates the phase information. Moreover, the likelihood analysis leads naturally to the matched filter which is the optimal filter for the network output. On the other hand, the coincidence strategy uses separate event lists formed by identifying candidate events in each detector considered in isolation - it ignores the crucial phase information that is inherent in a signal generated from a specific astrophysical source. This critical difference between the two analyses leads to the coherent analysis having superior detection performance. 
\par
Moreover, when we consider correlated noise between detectors, the coherent strategy again leads to matched filters which are optimal. The correlations between the noise enter naturally into the  likelihood analysis and the matched filter depends on these correlations. In coincidence analysis, on the other hand, the correlations between detectors simply do not play any role in the detection procedure and are thus \emph{ipso facto} ignored. 
\par
In this paper we have treated two detectors with identical noise PSD, identically oriented and in the same location. The results are easily generalized from two to $N$ detectors with identical noise PSD, with identical orientation and in the same location. We next propose to generalize these results to include arbitrarily oriented detectors in different geographical locations. We expect that the difference in performance between the two strategies to be more striking. One reason is that for differently oriented detectors, there is a finite possibility that a signal may not be detected separately by both detectors, thus ruling out coincidence detection, while coherently it could still be detected.
\par 
The work in this paper can be generalized in a straight forward manner to other GW sources with known waveforms of finite duration. More importantly, we believe that this work could be generalized to burst sources where the waveform could be constrained from physical and astrophysical considerations. Some work already exists in this direction \cite{SYL}. This may be a worthwhile future direction to adopt.           

\section{Acknowledgments}

S. Dhurandhar acknowledges the DST and JSPS Indo-Japan international cooperative programme for scientists and engineers for supporting visits to Osaka City University, Japan and Osaka National University, Japan. H. Tagoshi thanks JSPS and DST under the same Indo-Japan programme for his visit to IUCAA, Pune. H. Mukhopadhyay thanks CSIR for providing research scholarship and S. Mitra, IUCAA for helpful suggestions. 
This work was supported in part by Monbu Kagakusho Grant-in-aid for Scientific Research of Japan (Nos. 14047214, 16540251).
%%%%%%%%%%%%%%%%%%%%%%%%%%%%%%%%%%%%%%%%%%%%%%%%%%%%%%%%%%%%%%%%%%%%%%%%%%%

%%%%%%%%%%%%%%%%%%%%%%%%%%%%%%%%%%%%%%%%%%%%%%%%%%%%%%%%%%%%%%%%%%%%%%%%%%%

\end{document}